\documentclass[journal,onecolumn,11pt]{IEEEtran}
\usepackage{epsf, psfrag, amssymb, amsfonts, amsmath, cite, enumerate}
\usepackage{graphicx, subfigure, color, bbm, bm}
\usepackage{relsize}
\usepackage{exscale}
\usepackage{hyperref}   
\usepackage{mathtools}

\newtheorem{theorem}{Theorem}
\newtheorem{example}{Example}
\newtheorem{corollary}{Corollary}
\newtheorem{lemma}{Lemma}
\newtheorem{definition}{Definition}

\def\psfancypar#1#2{\begingroup\def\par{\endgraf\endgroup\lineskiplimit=0pt}
               \setbox2=\hbox{\large\sc #2}
               \newdimen\tmpht \tmpht \ht2 \advance\tmpht by \baselineskip
               \font\hhuge=Times-Bold at \tmpht
               \setbox1=\hbox{{\hhuge #1}}
               \count7=\tmpht \count8=\ht1
               \divide\count8 by 1000 \divide\count7 by \count8
               \tmpht=.001\tmpht\multiply\tmpht by \count7
               \font\hhuge=Times-Bold at \tmpht
               \setbox1=\hbox{{\hhuge #1}}
               \noindent
                \hangindent1.05\wd1
               \hangafter=-2 {\hskip-\hangindent
               \lower1\ht1\hbox{\raise1.0\ht2\copy1}%
                \kern-0\wd1}\copy2\lineskiplimit=-1000pt}

\newcommand{\beq}{\begin{equation}}
\newcommand{\eeq}{\end{equation}}
\newcommand{\bqa}{\begin{eqnarray}}
\newcommand{\eqa}{\end{eqnarray}}
\newcommand{\bqn}{\begin{eqnarray*}}
\newcommand{\eqn}{\end{eqnarray*}}
\newcommand{\nn}{\nonumber}

\newcommand{\be}{\begin{enumerate}}
\newcommand{\ee}{\end{enumerate}}
\newcommand{\bi}{\begin{itemize}}
\newcommand{\ei}{\end{itemize}}
\newcommand{\bd}{\begin{description}}
\newcommand{\ed}{\end{description}}
\newcommand{\ba}{\begin{array}}
\newcommand{\ea}{\end{array}}
\newcommand{\bde}{\begin{definition}}
\newcommand{\ede}{\end{definition}}
\newcommand{\bex}{\begin{example}}
\newcommand{\eex}{\end{example}}

\def\boxit#1{\vbox{\hrule\hbox{\vrule\kern3pt
        \vbox{\kern3pt#1\kern3pt}\kern3pt\vrule}\hrule}}

\def\reals{ { {\rm  I \kern-0.15em R }  } }
\def\complex{ {\,{{\rm C} \kern-0.50em \raise0.20ex {  |}}\, }}

\def\0bf{{\bf 0}}
\def\1bf{{\bf 1}}
\def\2bf{{\bf 2}}
\def\3bf{{\bf 3}}
\def\4bf{{\bf 4}}
\def\5bf{{\bf 5}}
\def\6bf{{\bf 6}}
\def\7bf{{\bf 7}}
\def\8bf{{\bf 8}}
\def\9bf{{\bf 9}}

\def\Rbf{{\bf R}}

\def\Rxx{\Rbf_{\ssstyle X\kern-.1em X}}

\let\ssstyle=\scriptscriptstyle

\def\Kout{\setbox1=\hbox{\Huge\bf K}\hbox to
1.05\wd1{\hspace{.05\wd1}
}}
\def\Sout{\setbox1=\hbox{\Huge\bf S}\hbox to 1.05\wd1{\hspace{.05\wd1}
}}

\newtheorem{remark}{Remark}
\newcommand*{\QEDA}{\hfill\ensuremath{\blacksquare}}%

\DeclareMathOperator*{\argmax}{arg\,max}

\begin{document}
\title{Asymptotically Optimal One- and Two-Sample Testing with Kernels}
\author{Shengyu Zhu, Biao Chen, Zhitang Chen, and  Pengfei Yang
	\thanks{This work was presented in part at the 22nd International Conference on Artificial Intelligence and Statistics (AISTATS), Naha, Okinawa, Japan, April 2019 \cite{Zhu2019}.}
    \thanks{Correspondence to: {zhushengyu@huawei.com}}
	}
\maketitle
\begin{abstract}
We characterize the asymptotic performance of nonparametric one- and two-sample testing. The exponential decay rate or  error exponent of the type-II error probability is used as the asymptotic performance metric, and an optimal test achieves the maximum rate subject to a constant level constraint on the type-I error probability. With Sanov's theorem, we derive a sufficient condition for one-sample tests to achieve the optimal error exponent in the universal setting, i.e., for any distribution defining the alternative hypothesis. We then show that two classes of Maximum Mean Discrepancy (MMD) based tests attain the optimal type-II error exponent on $\mathbb R^d$, while the quadratic-time Kernel Stein Discrepancy (KSD) based tests achieve this optimality with an asymptotic level constraint.  For general two-sample testing, however, Sanov's theorem is insufficient to obtain a similar sufficient condition. We proceed to establish an extended version of Sanov's theorem and derive an exact error exponent for the quadratic-time MMD based two-sample tests. The obtained error exponent is further shown to be optimal among all two-sample tests satisfying a given level constraint. {Our work hence provides an achievability result for optimal nonparametric one- and two-sample testing in the universal setting.} Application to off-line change detection and related issues are also discussed. 

\end{abstract}

\section{Introduction}
\label{sec:intro}
	
We study two  fundamental problems in statistical hypothesis testing: the one- and two-sample testing. One-sample testing, also referred to as goodness of fit testing, aims to determine how well a given distribution $P$ fits the observed sample $y^m\coloneqq \{y_i\}_{i=1}^m$. This goal can be achieved by testing  the null hypothesis $H_0: P=Q$ against the alternative hypothesis $H_1:P\neq Q$, where $Q$ is the true distribution governing the sample $y^m$. In two-sample or homogeneity testing, one wishes to test if two samples $x^n$ and $y^m$ originate from the same distribution. Let $P$ and $Q$ denote the underlying unknown distributions for the respective samples. Then a two-sample test  decides whether to accept  $H_0:P=Q$ or $H_1:P\neq Q$. 

Both one- and two-sample testing have a long history in statistics and find applications in a variety of areas. In anomaly detection \cite{li2014universal,Zou2017nonparametric,chandola2009anomaly}, the abnormal sample is supposed to come from a distribution that deviates from the typical distribution or sample. Similarly in change-point detection \cite{Basseville1993detectionchanges,Desobry2005onlinekernel,Harchaoui2009KernelChange,James1987testsforchange,Li2015Changedetection}, the post-change observations originate from a different source from the pre-change one. In bioinformatics, two-sample testing may be conducted to compare micro-array data from identical tissue types measured by different laboratories, to decide whether the data can be analyzed jointly \cite{Borgwardt2006}. Two-sample tests can also be applied to (conditional) independence testing by comparing the observed sample with a permutated version \cite{Gretton2012,Ramdas2015,doran2014permutationCI}, which is important to a class of methods, the so-called constraint-based methods, in causal discovery  \cite{spirtes2000causation}. Other examples include spectrum sensing in cognitive radio \cite{wang2009spectrum,denkovski2012hos}, criticizing statistical models\cite{Lloyd2015ModelCrit,Kim2016Crit}, and measuring  quality of samples drawn from a given probability density function (up to the normalization constant) by Markov Chain Monte Carlo (MCMC) methods \cite{Gorham2015Stein,Gorham2017measuringKSD,Chwialkowski2016Goodness}.

In this paper, we consider the {\it universal} nonparametric setting, in which no prior information on the unknown distributions is available. We will only allow for tests that are  independent of the unknown distributions, whereas the statistical performance of the tests may depend on the unknown distributions (cf.~Section~\ref{sec:problem}). Typical tests in this setting are constructed based on some probability distance measures between distributions, which possess the property that they are zero if and only if two distributions are identical; hence a larger sample estimate of the distance measure indicates that the two distributions are more likely to be different. Examples in some earlier tests include the Kolmogorov-Smirnov distance \cite{Kolmogorov1933,smirnov1948table,justel1997multivariate}, total variation distance \cite{Gyorfi1991}, and Wasserstein distance \cite{del1999tests,del2005asymptotics,ramdas2017wasserstein}. Although the constructed tests have satisfactory theoretic properties and work well in low dimensions (namely, $\mathbb R$), they do not in general apply to high dimension data. Recent tests have also used the Kullback-Leibler Divergence (KLD) \cite{Nguyen2010,kanamori2012f} and Sinkhorn divergence (smoothed Wasserstein distance) \cite{cuturi2013sinkhorn, ramdas2017wasserstein}, whose statistics are estimated by solving some optimization problems and can better handle higher dimensions. We refer the reader to related works, e.g., \cite{Gretton2012,Chwialkowski2016Goodness,ramdas2017wasserstein}, and references therein for a more detailed review of existing one- and two-sample tests.

More recently, kernel based statistics have attracted much attention in machine learning, as they possess several key advantages such as computational efficiency and fast convergence \cite{SmolaHilbert,Muandet2017KME}. A particular example is the Maximum Mean Discrepancy (MMD), defined by the distance between the mean embeddings of two distributions into a Reproducing Kernel Hilbert Space (RKHS) \cite{Gretton2012}.{\footnote{The MMD is closely related to another probability distance measure, the energy distance \cite{Baringhaus2004new,Szekely2004testing}. Roughly speaking, for every distance metric, there exists a suitable kernel (and also vice versa) so that the MMD and the energy distance are equivalent; see  \cite{Lyons2013distance,Sejdinovic2013equivalence} for details. In this paper, we focus on kernel based statistics, and particularly, the MMD and KSD.}} There have been several effective two-sample tests that are constructed based on the MMD: a vanilla test statistic can be computed by plugging in the sample empirical distributions with a quadratic-time computation complexity in terms of number of samples, and some variants have been proposed with even lower complexities \cite{Chwialkowski2015fast,Fukumizu2009,Gretton2009, Gretton2012OptKernelLarge,Sutherland2017GeneandCrit,Zaremba2013Btest}. Applying the MMD to one-sample testing is straightforward but requires integrals with respect to (w.r.t.) the target distribution $P$ \cite{Altun2006,Szabo2015Two, Szabo2016learning}. Another idea, in the context of model criticism, is to conduct a two-sample testing by drawing samples from $P$ \cite{Lloyd2015ModelCrit, Kim2016Crit}. A difficulty with this approach is to determine the required number of samples drawn from $P$ relative to $m$, the sample number of the test sequence. Alternatively, there exist more efficient one-sample tests constructed based on classes of Stein transformed RKHS functions \cite{Chwialkowski2016Goodness,Gorham2015Stein,Gorham2017measuringKSD,Liu2016GoodnessFit,Oates2017SteinRKHS},  where the test statistic is the norm of the smoothness-constrained function with largest expectation under $Q$ and is referred to as the Kernel Stein Discrepancy (KSD). The KSD has zero expectation under $P$ and does not require computing integrals or drawing samples. Additionally, constructing explicit features of distributions attains a linear-time one-sample test that is also more interpretable \cite{Jitkrittum2017linearGoodness}.


Distinguishing distributions with high success probability at a given fixed sample size, however, is not possible without any prior assumptions regarding the difference between $P$ and $Q$ (see an example for two-sample testing in \cite[Section~3]{Gretton2012}). Consequently, statistical performance in the universal setting are often considered in the large sample regime. A test is said to be consistent if its type-II error probability approaches zero in the limit, subject to a constant level constraint on the type-I error probability. While consistency is a desired property for hypothesis tests, it is even more desirable to characterize the decay rate w.r.t.~sample size as it provides a natural metric for comparing tests' performance. Indeed,  the decay rate of the type-II error probability has been investigated for existing kernel based one- and two-sample tests. For the one-sample tests in \cite{Altun2006,Szabo2015Two, Szabo2016learning} and two-sample tests in \cite{Chwialkowski2015fast,Fukumizu2009,Gretton2009, Gretton2012OptKernelLarge,Sutherland2017GeneandCrit,Zaremba2013Btest}, analysis is based on test statistics, through their asymptotic distributions or some probabilistic bounds on their convergence to the population statistics. The  statistical characterizations depend on kernels and are loose in general (more details are given in Section~\ref{sec:preliminary}). For KSD based one-sample tests, current characterization restricts to consistency; attempts in characterizing their asymptotic performance, in particular, the optimal error decay rate, have not been fruitful \cite{Chwialkowski2016Goodness, Jitkrittum2017linearGoodness,Liu2016GoodnessFit}. 


The present work is devoted to characterizing the statistical optimality of nonparametric one- and two-sample tests in the universal setting. This is motivated by the success of  Hoeffding's result in  \cite{Hoeffding1965} which established universal optimality of the so-called likelihood ratio test for testing against a multinomial distribution. For the general case, i.e., with arbitrary sample space, testing between $H_0: y^m\sim P$ and $H_1:y^m\sim Q$ can be extremely hard when $Q$ is arbitrary but unknown, as opposed to the simple case where $Q$ is known. With independent and identically distributed (i.i.d.)  sample and known $Q$, the type-II error probability of an optimal test, subject to a constant level constraint on the type-I error probability, vanishes exponentially fast w.r.t.~the sample size $m$, and the exponential decay rate or error exponent coincides with the KLD between $P$ and $Q$ (cf.~Lemma~\ref{lem:SteinLemma}). This motivates the so-called Universal Hypothesis Testing (UHT) problem \cite{Hoeffding1965}: {\emph{does there exist a nonparametric one-sample test that achieves the same optimal error exponent as in the simple hypothesis testing problem where $Q$ is known?}}{\footnote{Throughout the rest of this paper, the UHT problem refers to the specific problem of finding a universally optimal one-sample test in terms of the type-II error exponent.}}  Over the years, universally optimal tests are known to exist only when the sample space is finite \cite{Hoeffding1965,Unnikrishnan2011UHT}. For a more general sample space, attempts have been largely fruitless except the works of \cite{Tusnady1977asymptotically,Zeitouni1991universal,Yang2017RobustKLD}. {These results, however, were obtained under alternative criteria in the universal sense and the proposed tests were complicated for practical use with possibly high dimensional data (see Section~\ref{sec:problem1} for a detailed review)}. Here we remark that even the existence of an optimal test for the UHT problem remains unknown in the latter case. 

Closely related to the current setting is a broader class of composite hypothesis testing, where there is uncertainty in the distributions associated with the hypotheses. This uncertainty, if known {\it a priori}, could be used to devise tests to optimize the worst-case performance, leading to generalized likelihood  ratio tests or other minimax based tests, e.g., \cite{Feder2002universal}.  By contrast, the universal optimality criterion used in this paper is much stronger in that the optimum must be achieved for any distribution defining the alternative. Also related are the works \cite{Jitkrittum2017linearGoodness} and \cite{BalaMinimaxOptGOF}, which respectively use the approximate Bahadur slope  and detection boundary as performance metric to compare kernel based one-sample tests. The authors of \cite{Jitkrittum2017linearGoodness} show that their linear-time test has  a greater relative efficiency than the linear-time test proposed in \cite{Liu2016GoodnessFit}, assuming a mean-shift alternative. In \cite{BalaMinimaxOptGOF}, a nonparametric kernel based test is proposed to achieve the minimax optimality for a composite alternative. It is noted that, while the tests in \cite{Jitkrittum2017linearGoodness,BalaMinimaxOptGOF} are able to work in the universal nonparametric setting, the corresponding statisical results need to  assume a particular composite alternative.


\subsection{Contributions}
\label{sec:contributions} 
We show that a plug-in kernel test, comparing the MMD between the given distribution and the sample empirical distribution with a proper threshold, is an optimal approach to the UHT problem on Polish, locally compact Hausdorff space, e.g., $\mathbb R^d$. Taking into account the difficulty of obtaining closed-form integrals for non-Gaussian distributions, we then follow \cite{Lloyd2015ModelCrit} to cast one-sample testing into a two-sample problem. We establish the same optimality for the quadratic-time kernel two-sample tests proposed in \cite{Gretton2012}, provided that a suitable number of independent samples are drawn from the given distribution. For the KSD based tests, the constant level constraint on the type-I error probability is difficult to meet for all possible sample sizes. By relaxing the constraint to an asymptotic one, we show that the quadratic-time KSD based tests proposed in \cite{Chwialkowski2016Goodness,Liu2016GoodnessFit} are also optimal for the UHT problem under suitable conditions. Key to our approach are Sanov's theorem and the weak convergence properties of the MMD \cite{SimSch16Kernel,Sriperumbudur2016EstPM} and the KSD \cite{Gorham2017measuringKSD}, which enable us to directly investigate the acceptance region defined by the test, rather than using the test statistic as an intermediary. 

As another contribution, we investigate the quadratic-time kernel two-sample tests in a more general setting where the sample sizes scale in the same order. The original Sanov's theorem, however, is insufficient in this setting as it involves only a single distribution. To proceed, we derive an extended version of Sanov's theorem, based on which an exact type-II error exponent of the two-sample test is established. The obtained error exponent is then shown to be optimal among all two-sample tests under the same level constraint, and is independent of the choice of kernels  provided that they are bounded continuous and characteristic. 

Finally, we discuss related issues, including how two other statistical criteria, exact Bahadur slope and Chernoff index, perform under the universal optimality criterion. Application of our results to nonparametric off-line change detection is also included, and we establish an optimal change detection result when no prior information on the post-change distribution is available.

\subsection{Paper Organization}
Section~\ref{sec:problem} formally presents the problems of one- and two-sample testing, along with the optimality criterion used in this paper. A sufficient condition for a one-sample test to be universally optimal in terms of the type-II error exponent is  proposed in Section~\ref{sec:suff_cond}. We briefly review the MMD and KSD, and related tests in Section~\ref{sec:MMDKSD}.  Section~\ref{sec:one} presents two classes of MMD based tests and the KSD based tests that are optimal for the UHT problem. Section~\ref{sec:two} establishes an extended version of Sanov's theorem and shows that the quadratic-time MMD based two-sample test is also universally optimal for two-sample testing. We apply our results to nonparametric off-line change-point detection in Section~\ref{sec:app} and conclude this paper in  Section~\ref{sec:conclusion}.

Mostly standard notations are used throughout the paper. We use boldface $\mathbf P$ to denote probability of a set w.r.t.~a distribution specified by the subscript, e.g., $\mathbf P_{y^m\sim Q}(A)$ denotes the probability of $y^m\in A$ with $y^m$ i.i.d. $\sim Q$.
\section{Problem Statement} 
\label{sec:problem}
In this section, we formally state the problems of one- and two-sample testing and also introduce the optimality criterion used in this paper.
\subsection{One-Sample Testing}
\label{sec:problem1}
Throughout the rest of this paper, let $\mathcal X$ denote a Polish space (that is, a separable completely metrizable topological space) and $\mathcal P$ the set of Borel probability measures defined on $\mathcal X$. Given a distribution $P\in\mathcal P$ and a sample sequence $y^m$ from an unknown distribution $Q\in\mathcal P$, we want to determine whether to accept $H_0:P=Q$ or $H_1:P\neq Q$. A hypothesis test $\Omega(m)=\{\Omega_0(m),\Omega_1(m)\}$ partitions $\mathcal X^{m}$ into two disjoint sets with $\Omega_0(m)\cup\Omega_1(m)=\mathcal{X}^{m}$. If $y^m\in\Omega_i(m),i=0,1$, a decision is made in favor of hypothesis $H_i$. We say that $\Omega_0(m)$ is an acceptance region for the null hypothesis $H_0$ and $\Omega_1(m)$ the rejection region. A type-I error is made when $P=Q$ is rejected while $H_0$ is true, and a type-II error occurs when $P=Q$ is accepted despite $H_1$ being true. The two error probabilities are respectively 
\begin{align}
\alpha_m&\coloneqq \mathbf P_{y^m\sim P}\left(\Omega_1(m)\right),~\text{under}~H_0, \nn\\
\beta_m&\coloneqq \mathbf P_{y^m\sim Q}\left(\Omega_0(m)\right),~\text{under}~H_1.\nn
\end{align}
	
In general, the two error probabilities can not be minimized simultaneously. A commonly used approach is the Neyman-Pearson approach \cite{Casella2002} which imposes the type-I error probability constraint in the form of $\alpha_m\leq\alpha$ for a pre-defined $\alpha\in(0,1)$. A level $\alpha$ test is said to be consistent when the type-II error probability vanishes in the large sample limit. Such a test is exponentially consistent if the error probability vanishes exponentially fast w.r.t.~the sample size, i.e., when \[\liminf_{m\to\infty}-\frac{1}{m}\log \beta_m >0.\]
Here and  throughout the rest of this paper, $\log$ denotes the logarithm to the base $2$. The above limit is also referred to as the type-II error exponent \cite{Cover2006}.
	
We next present Chernoff-Stein lemma, which gives the optimal type-II error exponent of any level $\alpha$ test for simple hypothesis testing between two known distributions. Let $D(P\|Q)$ denote the KLD between $P$ and $Q$. That is, $D(P\|Q)=\mathbf E_P\log(dP/dQ)$ where $dP/dQ$ stands for the Radon-Nikodym derivative of $P$~w.r.t.~$Q$ when it exists, and  $D(P\|Q)=\infty$ otherwise \cite{Dembo2009}.
	
	\begin{lemma}[Chernoff-Stein Lemma {\cite{Cover2006, Dembo2009}}]
		\label{lem:SteinLemma}
		Let $y^m$ i.i.d.~$\sim R$. Consider hypothesis testing between $H_0: R = P\in\mathcal P$ and $H_1:R=Q\in\mathcal P$,
		with $0<D(P\|Q)<\infty$. Given $0<\alpha<1$, let $\Omega^*(m,P,Q)=(\Omega_0^*(m,P,Q), \Omega_1^*(m, P, Q))$ be the optimal level $\alpha$ test with which the type-II error probability is minimized for each sample size $m$. Then the type-II error probability satisfies 
		\[\lim_{m\to\infty}-\frac{1}{m}\log \mathbf P_{y^m\sim Q}(\Omega_0^*(m, P,Q)) = D(P\|Q).\]
	\end{lemma}

We can now describe the universal optimality criterion used in this paper. Let $\Omega(m)=(\Omega_0(m), \Omega_1(m))$ be a nonparametric one-sample test of level $\alpha$. With $y^m$ i.i.d.~$\sim Q$ under the alternative hypothesis $H_1$, the corresponding type-II error probability $Q(\Omega_0(m))$ can not be lower than the minimum  $Q(\Omega_0^*(m,P,Q))$. As a result, Chernoff-Stein lemma indicates that its type-II error exponent is upper bounded by $D(P\|Q)$. The UHT problem is  to find a test $\Omega(m)$ for a given $P$ so that
\begin{equation}
\label{uhtproblem}
\begin{aligned}
&\text{a) under}~H_0: \,\mathbf P_{y^m\sim P}(\Omega_1(m)) \leq \alpha, \\
&\text{b) under}~H_1: \,\liminf_{m\to\infty}-\frac1m \log \mathbf P_{y^m\sim Q}(\Omega_0(m))=D(P\|Q),
\end{aligned}
\end{equation}
for an arbitrary $Q$ with $0<D(P\|Q)<\infty$, giving rise to the name {\it universal} hypothesis testing. 

{It is also possible to consider a constant constraint on the type-I error exponent, i.e., for a given $\lambda>0$, the test $\Omega(m)$ needs to satisfy $\liminf_{m\to\infty}-\frac{1}{m}\log \mathbf P_{y^m\sim P}(\Omega_1(m))\geq\lambda$. Let $\hat Q_m$ denote the  empirical  measure  of sample $y^m$, i.e., $\hat Q_m=\frac{1}{m}\delta_{y_i}$ where $\delta_y$ is Dirac measure at $y$. When the sample space $\mathcal X$ is finite, Hoeffding's test is equivalent to deciding $H_0$ if $D(\hat Q_m\|P)\leq \lambda$ and $H_1$ otherwise, which is shown to achieve the optimal type-II error exponent among all tests that meet the type-I error exponent constraint \cite{Hoeffding1965}. For the problem considered in this paper, a properly vanishing threshold is needed to achieve the optimal type-II error exponent; in particular, Hoeffding's test is to decide $H_0$ if $D(\hat Q_m\|P)\leq \delta_m$ where $\delta_m\geq 0$ is such that the type-I error probability is equal to $\alpha$ and $\delta_m\to0$ as $m\to\infty$. It is important to note that although a constant constraint placed on the error exponent is more strict than that placed on the type-II error probability, an optimal test under the former constraint does not imply the optimality under the latter in general: one has to select a proper threshold, like $\delta_m$ in the above Hoeffding's test, to achieve the optimal type-II error exponent. 
}

{While Hoeffding's test has a desirable performance when $\mathcal X$ is finite, the construction of Hoeffding's test does not easily
extend to continuous sample spaces. In the following we provide a brief review on three existing attempts.

\begin{itemize}
\item
Authors of \cite{Zeitouni1991universal} considered a constant constraint placed on the type-I error exponent, i.e., the type-I error exponent is greater than or equal to a given $\lambda>0$. Let $d_L(\cdot,\cdot)$ denote the L{\'e}vy metric between two probability measures. For some $\delta>0$, denote $B_L(R, \delta)=\{S:d_L(R, S)<\delta\}$ as an open ball of radius $\delta$ around a distribution $R$ and $\Gamma^\delta=\cup_{R\in\Gamma}\{S\in\mathcal P: d_L(R,S)<\delta\}$ as the $\delta$-smooth set for a set $\Gamma\subset\mathcal P$. For $\Omega^*_1=\{S:\inf_{R\in B_L(S, 2\delta)}D(R\|P) > \lambda \}$ and $\Omega^*_0=\mathcal P\setminus\Omega_1^*$, it was shown that
    \begin{align}\liminf_{n\to\infty}-\frac1n\log \mathbf{P}_{y^m\sim P}(\hat Q_m\in{\Omega_1^*}^\delta)\geq\lambda, \nn
    \end{align}
and if there is another set $\Omega_1'$ and $\Omega_0'=\mathcal P\setminus \Omega_1'$ such that $\liminf_{n\to\infty}-\frac1n \mathbf P_{y^m\sim P}(\hat Q_m\in{\Omega'_1}^{6\delta})>\lambda$, then for any $Q\neq P$,
     \begin{align}\liminf_{n\to\infty}-\frac1n \mathbf P_{y^m\sim Q}(\hat Q_m\in{\Omega_0^*}^\delta)\geq\liminf_{n\to\infty}-\frac1n \log \mathbf{P}_{y^m\sim Q}(\hat Q_m\in{\Omega_0'}^{\delta}).
     \end{align} 
In this sense, the test $\Omega^*=(\Omega_0^*,\Omega_1^*)$ is said to be $\delta$-optimal. Although the above result holds for both finite and continuous sample spaces, the optimality is weaker than that of Hoeffding's test when the sample space is finite. Moreover, it is computationally challenging to decide whether an empirical measure $\hat Q_m$ falls into the set $\Omega_1^*$ due to the difficulty in practically characterizing the L{\'e}vy ball $B_L(S, 2\delta)$ as well as computing the infimum KLD over the ball. Besides, as discussed above,  this result does not necessarily indicate the optimality with a constant level constraint on the type-I error probability, i.e., when $\lambda=0$.


\item A difficulty of generalizing the Hoeffding's test from discrete sample space to continuous sample space is that the suplevel set $\{S\in\mathcal P: D(S\|P)\geq \lambda\}$ is not closed in $\mathcal P$, which leads to the unexpected property that its closure may encompass the entire probability space \cite{Hoeffding1965}. Thus, it was proposed in \cite{Yang2017RobustKLD} to consider a composite hypothesis testing by replacing $P$ with $B_L(P, \delta)$, i.e., $y^m$ is either from a distribution lying in $B_L(P, \delta)$ or another distribution $Q\notin B_L(P, \delta)$ but otherwise unknown. This is a different problem from ours. For this composite hypothesis testing problem, the test $\Omega_1^*=\{S: \inf_{R\in B_L(P, R)\leq\delta} D(S\|R) > \lambda\}$ and $\Omega_0^*=\mathcal P\setminus \Omega_1^*$ is shown to be optimal among all the tests that satisfy the type-I error exponent constraint with rate $\lambda$. However, this result was only established when the sample space is $\mathbb R$.
\item To extend Hoeffding's result to  $\mathbb R^d$, another work \cite{Tusnady1977asymptotically} proposed a test statistic based on the KLD between empirical measure and the one under $H_0$ on a finite partition of the sample space. A sufficient condition on the finite partition was then given to achieve optimal type-II error exponent with a constant constraint on both the type-I error probability and error exponent. Unfortunately, while such a condition can be used to verify if a partition suffices to achieve the optimality, it is not clear how to explicitly construct the partition for a given distribution $P$ and also when this condition is a necessary one.


\end{itemize} 
}
\subsection{Two-Sample Testing}
Let $x^n$ and $y^m$ be independent samples with $x^n\sim P$ and  $y^m\sim Q$, and  both $P$ and $Q$ are unknown. The goal of two-sample testing is to decide between $H_0:P=Q$ and $H_1:P\neq Q$ based on the observed samples. We use $\Omega(n,m)=(\Omega_0(n,m),\Omega_1(n,m))$ to denote a two-sample test, with $\Omega_0(n,m)\cap\Omega_1(n,m)=\varnothing$ and $\Omega_0(n,m)\cup\Omega_1(n,m)=\mathcal X^{n+m}$. The type-I and type-II error probabilities are given by
\begin{align}
	\alpha_{n,m} &\coloneqq  \mathbf P_{x^n\sim P,y^m\sim P}\left((x^n,y^m)\in \Omega_1(n,m)\right),~\text{under}~ H_0,\nn\\
	\beta_{n,m} &\coloneqq  \mathbf P_{x^n\sim P, y^m\sim Q}\left((x^n,y^m)\in\Omega_0(n,m)\right),~\text{under}~H_1,\nn
\end{align}
	respectively. Notice that  both $\alpha_{n,m}$ and $\beta_{n,m}$ are defined w.r.t.~the underlying yet unknown distributions under the respective hypotheses. 
	
Motivated by the UHT problem, we also consider the error exponent of $\beta_{n,m}$ defined in the large sample limit, with a constant level constraint on $\alpha_{n,m}$. That is, we would like to maximize
\begin{align}
\label{eqn:twotesting}
\liminf_{n,m\to\infty}-\frac{1}{n+m}\log\beta_{n,m},~\text{subject to}~\alpha_{n,m}\leq\alpha.
\end{align}
Unlike one-sample testing, there does not exist a characterization on the optimal type-II error exponent for two-sample testing. As such, we would like not only to derive an exact characterization of the type-II error exponent for a given two-sample test, but also to investigate if the characterization is optimal among all  two-sample tests satisfying the level constraint.

\section{A Sufficient Condition for Universal Hypothesis Testing}
\label{sec:suff_cond}
A useful tool for establishing the exponential decay of a hypothesis test is Sanov's theorem from large deviation theory. In this section, we will use it to derive a sufficient condition for one-sample tests to be universally optimal, followed by discussions on why various tests fail to meet this condition. 

We start with the weak convergence of probability measures, followed by Sanov's theorem. 

\begin{definition}[Weak Convergence]
For a sequence of probability measures $P_l\in\mathcal P$, we say that $P_l\to P$ weakly if and only if $\mathbf E_{x\sim P_l} f(x)\to\mathbf E_{x\sim P} f(x)$ for every bounded continuous function $f:\mathcal X\to\mathbb R$. The topology on $\mathcal P$ induced by this weak convergence is referred to as the weak topology.
\end{definition} 

\begin{theorem}[Sanov's Theorem {\cite{Sanov1958original,Dembo2009}}]
Let $y^m$ i.i.d.~$\sim Q\in\mathcal P$. Denote by $\hat Q_m$ the empirical measure of sample $y^m$, i.e., $\hat Q_m=\frac{1}{m}\delta_{y_i}$ where $\delta_y$ is Dirac measure at $y$. For a set $\Gamma\subset\mathcal P$ defined on the Polish space $\mathcal X$, we have
	\begin{align}
	\limsup_{m\to\infty}-\frac{1}{m}\log \mathbf P_{y^m\sim Q}(\hat Q_m\in\Gamma)&\leq\inf_{R\in\operatorname{int}\Gamma}D(R\|Q),\nn\\
	\liminf_{m\to\infty}-\frac{1}{m}\log\mathbf P_{y^m\sim Q}(\hat Q_m\in\Gamma)&\geq\inf_{R\in\operatorname{cl}\Gamma}D(R\|Q),\nn
	\end{align}
	where $\operatorname{int}\Gamma$ and  $\operatorname{cl}\Gamma$  are the interior and closure of $\Gamma$ w.r.t.~the weak topology, respectively.
\end{theorem}

A useful property of the KLD is its lower semi-continuity.
\begin{lemma}[Lower Semi-Continuity of the KLD {\cite{Hoeffding1965,VanErven2014Renyi}}]
\label{lem:lowerSemiContiKLD}
For a fixed $Q\in\mathcal P$, $D(\cdot\|Q)$ is a lower semi-continuous function w.r.t.~the weak topology of $\mathcal P$. That is, for any $\epsilon>0$, there exists a neighborhood $U\subset\mathcal P$ of $P$ such that for any $P'\in U$, $D(P'\|Q)\geq D(P\|Q)-\epsilon$ if $D(P\|Q)<\infty$, and $D(P'\|Q)\to\infty$ as $P'$ tends to $P$ if $D(P\|Q)=\infty$.
\end{lemma}
	
We can now present a sufficient condition which follows from Sanov's theorem and the lower semi-continuity of the KLD. 
\begin{theorem}
\label{thm:sufficient_UHT}
Let $y^m$~i.i.d.~$\sim Q$. Let $\Omega(m)=(\Omega_0(m),\Omega_1(m))$ be a one-sample test based on $y^m$ and $P$. Then it is optimal for the UHT problem if
\begin{enumerate}
    \item[a)] $\mathbf P_{y^m\sim P}(\Omega_1(m))\leq\alpha$ with $P=Q$.
    \item[b)] $\Omega_0(m)\subset\{y^m:d(P, \hat Q_m)\leq \gamma_m\}$, where $d(\cdot,\cdot)$ is a probability metric that metrizes the weak topology on $\mathcal P$, $\hat Q_m$ denotes the empirical measure of $y^m$, $\gamma_m>0$ denotes the test threshold and goes to $0$ as $m\to\infty$.
\end{enumerate}
\end{theorem}
\begin{IEEEproof} 
Condition~a) is simply the constant constraint on the type-I error probability.  By Chernoff-Stein lemma, we only need to show that the type-II error exponent is no lower than $D(P\|Q)$. Assuming Condition~b), we have 
\begin{align}
\label{eqn:proof_suff}
\liminf_{m\to\infty} -\frac{1}{m}\log \mathbf P_{y^m\sim Q}(\Omega_0(m))\geq\liminf_{m\to\infty} -\frac{1}{m}\log \mathbf P_{y^m\sim Q}\left (d(P,\hat Q_m)\leq \gamma_m\right).
\end{align}
To proceed, we notice that deciding if $y^m\in\{y^m:d(P,\hat Q_m)\leq \gamma_m\}$ is equivalent to deciding if the empirical measure $\hat Q_m\in \{P':d(P,P')\leq \gamma_m\}$. Since $\gamma_m\to0$ as $m\to\infty$, for any given $\gamma>0$, there exists an integer $m_0$ such that $\gamma_m<\gamma$ for all $m>m_0$. Therefore, $\{P':d(P,P')\leq \gamma_m\}\subset\{P':d(P,P')\leq\gamma\}$ for large enough $m$, and for any $\gamma>0$,
	\begin{align}
	\label{eqn:common1}
&\,\liminf_{m\to\infty}-\frac{1}{m}\log \mathbf P_{y^m\sim Q}\left(d(P,\hat Q_m)\leq\gamma_m \right)\nn\\\geq&\,\liminf_{m\to\infty}-\frac{1}{m}\log \mathbf P_{y^m\sim Q}\left(d(P,\hat Q_m)
		\leq\gamma \right)\nn\\
		\geq&\,\inf_{\{P'\in\mathcal P:d(P,P')\leq\gamma\}} D(P'\|Q),
		\end{align}
		where the last inequality is from Sanov's theorem and that $\{P':d(P,P')\leq\gamma\}$ is closed w.r.t.~the weak topology. Moreover, for any given $\epsilon>0$, there exists some $\gamma>0$ such that 
		\begin{align}
		\label{eqn:common2}
\inf_{\{P':d_k(P,P')\leq\gamma\}} D(P'\|Q)\geq D(P\|Q)-\epsilon,
		\end{align}
		 due to the lower semi-continuity of the KLD in Lemma~\ref{lem:lowerSemiContiKLD} and the assumption that $0<D(P\|Q)<\infty$ under $H_1$. Since $\epsilon$ can be arbitrarily small, combining (\ref{eqn:proof_suff}), (\ref{eqn:common1}) and (\ref{eqn:common2}) gives
		\begin{align}
		\liminf_{m\to\infty}-\frac{1}{m}\log \mathbf P_{y^m\sim Q}(\Omega_0(m))\geq D(P\|Q),~\text{under}~H_1: P\neq Q,\nn
		\end{align}
which completes the proof.
\end{IEEEproof}
\begin{remark}
It is worth noting  that Condition b) requires only a vanishing threshold $\gamma_m$ and does not place any constraint on how fast it vanishes. Indeed, the requirement on the vanishing rate is determined by the type-I error constraint in Condition a). Consequently, if a test has its acceptance region in the form of $\{y^m: d(P,\hat Q_m)\leq\gamma_m\}$ and is universally optimal, then any such test with a vanishing threshold $\gamma_m'>0$, where $\gamma_{m}'\geq\gamma_m$, also satisfies the two conditions, as $\{y^m: d(P,\hat Q_m)\leq\gamma_m\}\subset\{y^m: d(P,\hat Q_m)\leq\gamma'_m\}$. However, using a larger threshold may result in a higher type-II error probability in the finite sample regime. Several methods have been proposed to choose a tighter threshold but they introduce additional randomness. More discussions will be given in Section~\ref{sec:remark}. 
\end{remark}
\begin{remark}
\label{rmk:generaltwosample}
A direct extension of the above result is to obtain a similar theorem for nonparametric two-sample tests.  However, Sanov's theorem works only with a single distribution, whereas there are two distributions involved in two-sample testing. Extending Sanov's theorem to handle two distributions would be key to establishing a sufficient condition for two-sample testing similar to Theorem~\ref{thm:sufficient_UHT}. This is the topic of Section~\ref{sec:two_Sanov} . 
\end{remark}

While Theorem~\ref{thm:sufficient_UHT} is somewhat straightforward since most of the hard work has  been done in proving Sanov's theorem {\cite{Sanov1958original,Dembo2009}, the two conditions are indeed quite hard to meet simultaneously.  For example, the KLD and total variation distance do not metrize weak convergence and  tests that are constructed from them fail to meet Condition b). While other distances, such as L{\'e}vy metric, Wasserstein distance, and the bounded Lipstchiz metric, metrize weak convergence, their sample estimates are usually not easy to compute. Moreover, there does not exist a uniform threshold such that  Condition a) is satisfied. To the best of our knowledge, universally optimal one-sample tests in the sense of (\ref{uhtproblem}) only exist for distributions defined on a finite sample space where the empirical KLD \cite{Hoeffding1965} or mismatched distance \cite{Unnikrishnan2011UHT} are used for constructing one-sample tests. Clearly, seeking a proper probability distance becomes key to meeting the sufficient condition given in Theorem~\ref{thm:sufficient_UHT}.  

    
    

 Meanwhile, in the machine learning community, there has been an active research topic on kernel based probability distances. While several efficient tests have been constructed based on these probability distances, little is known about their statistical optimality. In the next section, we will introduce two such kernel based probability distances and their empirical estimates for constructing nonparametric one- and two-sample tests.


\section{Maximum Mean Discrepancy and Kernel Stein Discrepancy} 
\label{sec:MMDKSD}
We introduce two kernel based probability distances, followed by a brief review of related one- and two-sample tests.
\subsection{Maximum Mean Discrepancy}
\label{sec:MMD}

{Let $\mathcal H$ denote a Hilbert space (a complete, possibly infinite-dimensional vector space endowed
with an inner product)  of real-valued functions on $\mathcal X$, where $\langle\cdot, \cdot\rangle_{\mathcal H}:\mathcal H\times\mathcal H \to\mathbb R$ denotes the inner product and $\|\cdot\|_{\mathcal H}:\mathcal H\to\mathbb R$ is the associated norm defined as $\|f\|_{\mathcal H}=\sqrt{\langle f, f\rangle_{\mathcal H}}$ for $f\in\mathcal H$. The evaluation functional over $\mathcal H$ is a linear functional $\mathcal F_x$ that evaluates each function at a point $x\in\mathcal X$, i.e.,  $\mathcal F_x[f] = f(x)$ for all $f\in\mathcal H$. Then $\mathcal H$ is a Reproducing Kernel Hilbert Space (RKHS) if the evaluation functional is bounded, i.e., for all $x\in\mathcal X$, there exists some finite $C>0$ such that $|\mathcal F_x(f)|=|f(x)|\leq C\|f\|_{\mathcal{H}}$ for all $f\in\mathcal H$. 

Given an RKHS, there exists a function $k_x\in\mathcal H$ such that $\mathcal F_x[f]=\langle k_x, f\rangle_{\mathcal H} = f(x)$  for each $x\in\mathcal X$, according to the Riesz representation theorem \cite{Akhiezer1993}. The corresponding reproducing kernel $k:\mathcal X\times\mathcal X\to\mathbb R$ is a function defined by $k(x,y)=k_y(x)$, which is positive definite on $\mathcal X\times\mathcal X$ in the sense that $\sum_{i=1}^n\sum_{j=1}^nc_ic_jk(x_i,x_j)\geq 0$ for any positive integer $n$, any choice of $x_1,\cdots, x_n\in\mathcal X$, and any $c_1, \cdots, c_n\in\mathbb R$. In \cite{aronszajn1950theory}, it was shown that for each positive definite function $k$ there exists a unique RKHS with $k$ as its reproducing kernel; conversely, the reproducing kernel of an RKHS is unique and positive definite. Thus, we can focus on the specific kernel associated with an RHKS and will use $\mathcal H_k$ to denote the RKHS to make the kernel explicit. 
}

Now let $x$ be an $\mathcal X$-valued random variable with probability measure $P$, and $\mathbf E_{x\sim P}f(x)$ the expectation of $f(x)$ for a function $f:\mathcal X\to\mathbb R$. Assume that $k$ is bounded continuous. Then for every Borel probability measure $P$ defined on $\mathcal X$, there exists a unique element $\mu_k(P)\in\mathcal H_k$ such that $\mathbf E_{x\sim P}f(x)=\langle f, \mu_k(P)\rangle_{\mathcal H_k}$ for all $f\in\mathcal H_k$ \cite{Berlinet2011RKHS}. The MMD between two Borel probability measures $P$ and $Q$ is defined as 
\[d_k(P,Q) \coloneqq  \sup_{\|f\|_{\mathcal H_k}\leq1}\left(\mathbf E_{x\sim P} f(x)-\mathbf E_{x\sim Q} f(x)\right),\]
where $\|f\|_{\mathcal{H}_k}\leq1$ denotes the unit ball in the RKHS \cite{Gretton2012}. The MMD belongs to a class of metrics for probability measures, called the Integral Probability Metric (IPM). Choosing an appropriate set of functions over which the supremum is taken can obtain many other popular distance measures, including the total variation distance and Wasserstein distance. We refer the reader to \cite{Sriperumbudur2012empirical} for more details on the IPM.  

The use of the unit ball in the RKHS brings in an equivalent formulation of the MMD. It is shown in \cite{Gretton2012} that the MMD can also be expressed as the RKHS-distance between $\mu_k(P)$ and $\mu_k(Q)$:
	\begin{align}
	\label{eqn:defintionMMD}
	d_k(P,Q)=&\,\|\mu_k(P)-\mu_k(Q)\|_{\mathcal H_k}=\big(\mathbf E_{x,x'}k(x,x')+\mathbf{E}_{y,y'}k(y,y')-2\mathbf E_{x,y}k(x,y)\big)^{1/2},
	\end{align}
	where $x,x'$ i.i.d.~$\sim P$ and $y,y'$ i.i.d.~$\sim Q$. {A kernel $k$ is said to be {characteristic} if the the map $u_k:\mathcal P\to \mathcal H_k$ is injective. Intuitively, this indicates that the RKHS endowed with the kernel $k$ should contain a sufficiently rich class of functions to represent all higher order moments of $P$.  There exist several characterizations for kernels to be characteristic and  a summary can be found in \cite[Section~3.3]{Muandet2017KME}}. With a characteristic kernel $k$, the MMD $d_k(\cdot, \cdot)$ becomes a metric on $\mathcal P$ \cite{Gretton2012,Sriperumbudur2010hilbert}, which enables the MMD to distinguish between different distributions. Moreover, \cite{SimSch16Kernel,Sriperumbudur2016EstPM} have also established the weak metrizable property of $d_k(\cdot,\cdot)$, as stated below.\footnote{Indeed, it is shown in  \cite{SimSch16Kernel} that $\mathcal X$ only needs to be locally compact Hausdorff. We require $\mathcal X$ be Polish in order to utilize Sanov's theorem.}
	
\begin{theorem}[\hspace{-0.01pt}{\cite{SimSch16Kernel, Sriperumbudur2016EstPM}}]
		\label{thm:MMDmetrize}
		The MMD $d_k(\cdot,\cdot)$ metrizes the weak convergence on $\mathcal P$ if the following conditions hold:
		\begin{itemize}
		\item({\bf A1}) the sample space $\mathcal X$ is Polish, locally compact and Hausdorff;
		\item({\bf A2}) the kernel $k$ is bounded continuous and characteristic.
		\end{itemize}
\end{theorem}	


As discussed in Section~\ref{sec:suff_cond}, the weak metrizable property is key to the sufficient condition in Theorem~\ref{thm:sufficient_UHT}. We note that this property is also important to training deep generative models \cite{Arjovsky2017WGAN,Li2017mmdGAN} in machine learning. Examples under Condition \hyperref[thm:MMDmetrize]{\bf{A1}} include any finite set and $\mathbb R^d$, and Condition \hyperref[thm:MMDmetrize]{\bf A2} is satisfied by Gaussian and Laplace kernels defined on $\mathbb R^d$, which are 
\begin{align}
\label{eqn:kernels}
k(x,y) = e^{-\|x-y\|^2/\gamma}~\text{and}~k(x,y)=e^{-\|x-y\|/\gamma},
\end{align}
respectively, with $x,y\in\mathbb R^d$ and $\gamma>0$ being a kernel parameter. {The Gaussian kernel is widely used in kernel related methods such as kernel mean embeddings and support vector machines, for its ease of computation (see Examples \ref{exp1}--\ref{exp3}) as well as its universality---a kernel is universal if the corresponding RKHS is dense in the space of bounded continuous functions defined on a compact metric space \cite{Micchelli2006universal}.}

\subsection{Kernel Stein Discrepancy}
The KSD is recently proposed in \cite{Chwialkowski2016Goodness,Liu2016GoodnessFit} and calculates some discrepancy from distribution $Q$ to a given distribution $P$ by using a Stein operator. {For the rest of this paper where KSD is involved, we assume that the sample space $\mathcal X$ is $\mathbb R^d$ or a compact subset of $\mathbb R^d$, and that the density functions (w.r.t.~Lebesgue measure) of $P$ and $Q$ exist and are continuously differentiable, which are denoted as $p$ and $q$, respectively.}

The KSD is defined as 
\begin{align}
d_S(P,Q)\coloneqq \sup_{\|f\|_{\mathcal H_k}\leq 1} \mathbf E_{x\sim Q}\left(s_p(x)f(x)+\nabla_xf(x)\right),\nn
\end{align}
where the function space $\|f\|_{\mathcal H_k}\leq 1$ denotes the unit ball of an RKHS $\mathcal H_k$, and $s_p(x)=\nabla_x\log p(x)$ is the score function of $p(x)$. The above definition requires $f$ be differentiable. Fortunately, according to \cite[Corollary~4.36]{steinwart2008support}, any function $f\in\mathcal H_k$ is $t$-times differentiable if the kernel $k$ is $t$-times differentiable. Hence we only need to pick a proper kernel $k$ for $d_S(P,Q)$ to be well-defined.} In particular, with a $\mathcal C_0$-universal kernel\footnote{`$\mathcal C_0$-universal' generalizes `universal' defined over a compact metric space to a non-compact one like $\mathbb R$. A kernel is $\mathcal C_0$-universal if the corresponding RKHS $\mathcal H_k$ is dense in the Banach space of continuous functions vanishing at infinity with the uniform norm.} \cite[Definition~1 and Theorem~4.1]{Carmeli2010vector} and $\mathbf E_{x\sim Q}\|\nabla_x\log p(x)-\nabla_x\log q(x)\|^2\leq\infty$,  $d_S(P,Q)=0$ if and only if $P=Q$ \cite[Theorem~2.2]{Chwialkowski2016Goodness}. A nice property of the KSD is that this result requires only the knowledge of $p(x)$ up to a normalization constant. To see this, notice that $\nabla_x\log p(x) = \nabla_x\log (\eta p(x))$ for any constant $\eta>0$. An equivalent expression of the squared KSD can be derived as
 \[d_S^2(P,Q)=\mathbf E_{x\sim Q}\mathbf E_{x'\sim Q}\, h_p(x,x'),\]
where \begin{align}
\label{eqn:KSD_h}
h_p(x,x')\coloneqq s_p^T(x)s_p(x')k(x,x')+s_p^T(x')\nabla_xk(x,x')+s_p^T(x)\nabla_{x'}k(x,x')+\operatorname{trace}(\nabla_{x,x'}k(x,x')).
\end{align}

Although the KSD is not a probability metric on $\mathcal P$, it has been shown to be  lower bounded in terms of some weak metrizable measures under proper conditions, as stated below. 
\begin{theorem}[\hspace{0.01pt}\cite{Gorham2017measuringKSD}]
\label{thm:KSD_weak}
If a) $\mathcal X=\mathbb R$ and $k(x,y)=\Phi(x-y)$ for some $\Phi\in{\mathcal C^2}$ (twice continuous differentiable) with a non-vanishing generalized Fourier transform; b) $k(x,y)=\Phi(x-y)$ for some $\Phi\in{C^2}$ with a non-vanishing generalized Fourier transform and the sequence $\{\hat Q_m\}_{m\geq1}$ is uniformly tight, then there exists a weak metrizable measure $d_W$ such that
\[d_W(P,\hat Q_m)\leq g(d_S(P, \hat Q_m)),\]
where $g$ is a function involving some unknown constants and $g(w)\to0$ if $w\to0$.
\end{theorem} 

This theorem indicates that $d_S(P,\hat Q_m)\to 0$ only if $\hat Q_m\to P$ weakly, i.e., $d_S(P,\hat{Q}_m)$ vanishing is a necessary condition for weak convergence. The Gaussian kernel defined on $\mathbb R$ satisfies Condition~a), and an example under Condition b) is the inverse multi-quadric kernel \[k(x,y)=(c^2+\|x-y\|^2)^\eta,~\text{with}~c>0~\text{and}~-1<\eta<0.\]

\subsection{Preliminary Results}
\label{sec:preliminary}
We end this section with some preliminary results on MMD and KSD based one- and two-sample tests. As one will see, these results depend on kernels and are generally loose. Nevertheless, they are important  to finding a proper test threshold to meet  the level constraint on the type-I error probability.

\subsubsection{MMD based one-sample test statistic} From the definition of the MMD, a one-sample test statistic  can be directly obtained by plugging in the empirical distribution of the observed sample. With sample $y^m$ and its empirical distribution $\hat Q_m$, the squared MMD can be estimated as
\begin{equation}
\label{eqn:onesampleMMD}
d_k^2(P, \hat Q_m) =\mathbf{E}_{x,x'}k(x,x') +  \frac{1}{m^2}\sum_{i=1}^m\sum_{j=1}^mk(y_i,y_j)
-\frac{2}{m}\sum_{i=1}^m\mathbf E_xk(x,y_i),
\end{equation}
where $x,x'$~i.i.d.~$\sim P$. A statistical characterization on this statistic is given as follows.
\begin{lemma}[\hspace{-0.01pt}\cite{Altun2006, Szabo2015Two}]
	\label{lem:gamman}
	Assume \hyperref[thm:MMDmetrize]{\bf A1} and \hyperref[thm:MMDmetrize]{\bf A2}, with $0\leq k(\cdot,\cdot)\leq K$. Given $y^m~\text{i.i.d.}\sim Q$, we have
	\[\mathbf{P}_{y^m\sim Q}\left(\left|d_k(P, \hat{Q}_{m})-d_k(P,Q)\right|>\left(2{K}/{m}\right)^{1/2}+\epsilon\right)\leq \exp{\left(-\frac{\epsilon^2m}{2K}\right)}.\]
\end{lemma}
\subsubsection{MMD based two-sample test statistic} Given two samples $x^n$ and $y^m$, a two-sample test statistic can be constructed as
\begin{equation}
\label{eqn:twosampleMMD}
d_k^2(\hat P_n,\hat Q_m)=\frac{1}{n^2}\sum_{i=1}^n\sum_{j=1}^nk(x_i,x_j)+\frac{1}{m^2}\sum_{i=1}^m\sum_{j=1}^mk(y_i,y_j)-\frac{2}{nm}\sum_{i=1}^n\sum_{j=1}^mk(x_i,y_j).
\end{equation}
where $\hat P_n$ and $\hat Q_m$ are the empirical distributions of $x^n$ and $y^m$, respectively. This statistic was proposed in \cite{Gretton2012} and is a biased estimator of $d_k^2(P,Q)$. The following lemma states the convergence of $d_k(\hat P_n, \hat Q_m)$ to $d_k(P,Q)$
\begin{lemma}[\hspace{-0.01pt}{\cite[Theorem 7]{Gretton2012}}]
	\label{lem:gammanm}	
 Assume the same conditions in Lemma~\ref{lem:gamman}. With $x^n$ i.i.d. $\sim P$ and $y^m$ i.i.d. $\sim Q$, we get
	\[\mathbf{P}_{x^n\sim P, y^m\sim Q}\left(\big|d_k(\hat P_n,\hat Q_m)-d_k(P,Q)\big|>2(K/n)^{{1}/{2}}+2(K/m)^{{1}/{2}}+\epsilon\right)
	\leq 2\exp\left(-\frac{\epsilon^2nm}{2K(n+m)}\right).\]
\end{lemma} 

\subsubsection{KSD based one-sample test statistic} Given sample $y^m$, we may estimate $d_S^2(P, Q)$ by 
\[d_S^2(P,\hat Q_m)=\frac{1}{m^{2}}\sum_{i=1}^m \sum_{j=1}^m h_p(y_i,y_j),\] where $h_p(\cdot,\cdot)$ is defined in (\ref{eqn:KSD_h}). The statistic $d_S^2(P,\hat Q)$ is a degenerate V-statistic under the null hypothesis $H_0:P=Q$ \cite{Chwialkowski2016Goodness}. To our best knowledge, existing results only characterize its limiting behavior, as stated in the following lemma.\footnote{The authors of \cite{Chwialkowski2016Goodness} assume $\tau$-mixing as the notion of dependence within the observations, which holds in the i.i.d.~case. They also assume a technical condition $\sum_{t=1}^\infty t^2\sqrt{\tau(t)}\leq\infty$ on $\tau$-mixing. See details in \cite{Chwialkowski2016Goodness, Dedecker2007}.} 
\begin{lemma}[\hspace{-0.01pt}{\cite[Proposition~3.1]{Chwialkowski2016Goodness}}]
\label{lem:KSDnull}
Assume the density functions $p$ and $q$ are continuously differentiable and kernel $k$ is also differentiable over $\mathcal X$. If $h_p$ is Lipschitz continuous and $\mathbf E_{x\sim Q} h_p(x,x)<\infty$, then  $md_S^2(P, \hat Q_m)$ converges weakly to some distribution under the null hypothesis $H_0$.
\end{lemma}

Letting $P=Q$ so that $d_k(P,Q)=0$ in Lemmas~\ref{lem:gamman} and \ref{lem:gammanm}, one can easily obtain a distribution-free threshold to meet the constant type-I error constraint. With such a threshold, however, the type-II error probability under the alternative hypothesis $H_1:P\neq Q$ depends on the kernel $k$ (more precisely, an upper bound $K$) and the resulting type-II error exponent is not tight. As to the KSD based test statistic, since there is no finite sample result like Lemmas~\ref{lem:gamman} and \ref{lem:gammanm}, the constant level constraint can not be satisfied for each sample size $m$.  In Section~\ref{sec:KSD1sampletest}, we will relax the level constraint to an asymptotic one for KSD based one-sample tests.


\section{Asymptotically Optimal One-Sample Tests}
\label{sec:one}
In this section, we investigate three classes of kernel based one-sample tests for the  UHT problem: the first test directly computes the MMD between the given distribution and the sample empirical distribution, which requires  closed-form integrals w.r.t.~the given distribution; the second test relaxes the exact integration by drawing samples from the target distribution but needs more treatment in applying Sanov's theorem; and the third test is more computationally favorable but only meets an asymptotic level constraint.

\subsection{Plug-in Kernel Tests}
The first test relies on the statistic $d_k(P, \hat Q_m)$ defined in (\ref{eqn:onesampleMMD}) and has been studied in  \cite{Altun2006,BalaMinimaxOptGOF,Szabo2015Two, Szabo2016learning}, yet its optimality for the UHT problem remains unknown.
	
Let  $\hat Q_m$ be the empirical measure of $y^m$. A plug-in kernel test can be constructed with acceptance region
\[\Omega_0(m)= \left\{y^m:d_k(P, \hat Q_m)\leq\gamma_{m}\right\},\]	where 
 $\gamma_m$ denotes the test threshold. On the one hand, we want the threshold $\gamma_m$ to be small so that the test type-II error probability is low; on the other hand, the threshold cannot be too small in order to satisfy the level constraint on the type-I error probability. The balance between the two error probabilities is attained with a  threshold that vanishes at an appropriate rate. 
\begin{theorem}
\label{thm:simple1}
For $P\in\mathcal P$ and $y^m$ i.i.d.~$\sim Q\in\mathcal P$, assume $0<D(P\|Q)<\infty$ under the alternative hypothesis $H_1$.	Assume \hyperref[thm:MMDmetrize]{\bf A1}, \hyperref[thm:MMDmetrize]{\bf A2}, where kernel $k$ satisfies $0\leq k(\cdot,\cdot)\leq K$ and $K>0$ is a constant value. For a given~$\alpha$, $0<\alpha<1$, set $\gamma_m=\sqrt{2K/m}\left(1+\sqrt{-\log\alpha}\right).$ Then the kernel test $d_k(P,\hat Q_n)\leq \gamma_m$ is an optimal level $\alpha$ test for the UHT problem, that is, 
\begin{align}
&\text{a) under}~H_0:\,\mathbf P_{y^m\sim P}\left(d_k(P,\hat Q_m)>\gamma_m\right)\leq\alpha,\nn\\
&\text{b) under}~H_1:\,\liminf_{m\to\infty}-\frac1m \log \mathbf P_{y^m\sim Q} \left(d_k(P,\hat Q_m)\leq\gamma_m\right)=D(P\|Q).\nn
\end{align}
\end{theorem}
\begin{IEEEproof}
That the test $d_k(P,\hat Q_m)\leq\gamma_m$ has level $\alpha$ can be directly verified by Lemma~\ref{lem:gamman} with $P=Q$. The rest follows  from Theorem~\ref{thm:sufficient_UHT}, since $d_k(\cdot, \cdot)$ metrizes weak convergence on $\mathcal P$ and $\gamma_m\to0$ as $m\to 0$.
\end{IEEEproof}
{
\begin{example}
\label{exp1}Consider a one-sample problem where $\mathcal X=\mathbb R$ and $P$ follows  standard Gaussian distribution $\mathcal N(0,1)$. We use the Gaussian kernel given in (\ref{eqn:kernels}) and pick $\gamma=2$, i.e., $k(x,y) =\exp(-(x-y)^2/2)$. An upper bound on this kernel function is  $K=1$ and the threshold becomes $\gamma_m=\sqrt{2/m}(1+\sqrt{-\log\alpha})$. For the test statistic defined in (\ref{eqn:onesampleMMD}), we have
\begin{align}
    d_k^2(P, \hat Q_m) =&~ \int_{-\infty}^{\infty}\int_{-\infty}^{\infty}e^{-(x-x')^2/2}\frac{1}{\sqrt{2\pi}}e^{-x^2/2}\frac{1}{\sqrt{2\pi}}e^{-x'^2/2}dxdx'+\frac{1}{m^2}\sum_{i=1}^m\sum_{j=1}^m e^{-(y_i-y_j)^2/2}\nn\\
    &~-\frac{2}{m}\sum_{i=1}^m\int_{-\infty}^{\infty}e^{-(x-y_i)^2/2}\frac{1}{\sqrt{2\pi}}e^{-x^2/2}{dx} \nn\\
    =&~{\frac{\sqrt{3}}{3}}+\frac{1}{m^2}\sum_{i=1}^m\sum_{j=1}^m e^{-(y_i-y_j)^2/2}-\frac{\sqrt{2}}{m}\sum_{i=1}^m e^{-y_i^2/4},\nn
\end{align}
whose evaluation requires $\mathcal O(m^2)$ time complexity.

\end{example}
}

The statistic $d_k^2(P,\hat Q_m)$ is a biased estimator for $d_k^2(P,Q)$. Replacing $\frac{1}{m^2}\sum_{i=1}^m\sum_{j=1}^m k(y_i,y_j)$ in $d_k^2(P,\hat Q_m)$ with $\frac{1}{m(m-1)}\sum_{i=1}^m\sum_{j\neq i}k(y_i,y_j)$ results in an unbiased statistic, which is denoted as $d_u^2(P,\hat Q_n)$. We remark that $d_u^2(P, \hat Q_m)$ is not a squared quantity and can be negative, a consequence of its unbiasedness. The following result shows that $d_u^2(P,\hat Q_m)$ can also be used to construct a universally optimal one-sample test.
	
	\begin{corollary}
		\label{cor:simple2}
		Under the same conditions as in Theorem~\ref{thm:simple1}, the test $d_u^2(P,\hat Q_m)\leq\gamma_m^2+K/m$ is level $\alpha$ and optimal  for the UHT problem.
	\end{corollary}
	\begin{IEEEproof}
	Since $0\leq k(\cdot,\cdot)\leq K$, we have 
\begin{align}
\left|d_u^2(P,\hat Q_m)-d^2_k(P,\hat Q_m)\right|=\left|\frac1{m^2(m-1)}\sum_{i=1}^m\sum_{j\neq i}k(x_i,x_j)-\frac1{m^2}\sum_{i=1}^mk(x_i,x_i)\right|
\leq{K}/{m}.\nn
\end{align} 
Thus, 
\begin{align}
\left\{y^m:d_k^2(P,\hat Q_m)\leq\gamma_m^2\right\}\subset\left\{y^m:d_u^2(P,\hat Q_m)\leq\gamma_m^2+K/m\right\}\subset\left\{y^m:d_k^2(P,\hat Q_m)\leq\gamma_m^2+2K/m\right\}.\nn
\end{align}
Under $H_0$, we  have 
\begin{align}
\mathbf P_{y^m\sim P}\left(d_u^2(P,\hat Q_m)>\gamma_m^2+K/m\right)\leq \mathbf P_{y^m\sim P}\left(d_k^2(P,\hat Q_m)>\gamma_m^2\right)\leq\alpha,\nn
\end{align}
where the last inequality is from Lemma~\ref{lem:gamman} and the fact that $d_k(P,\hat Q_m)\geq 0$. Under $H_1:P\neq Q$, we have the type-II error exponent being
\begin{align}
&\,\liminf_{m\to\infty}-\frac{1}n\log \mathbf P_{y^m\sim Q}\left(d_u^2(P,\hat Q_m)\leq\gamma_m^2+K/m\right)\nn\\
\geq&\,\liminf_{n\to\infty}-\frac{1}m\log \mathbf P_{y^m\sim Q}\left(d_k^2(P,\hat Q_m)\leq\gamma_m^2+2K/m\right)\nn\\
=&\,\liminf_{n\to\infty}-\frac{1}m\log \mathbf P_{y^m\sim Q}\left(d_k(P,\hat Q_m)\leq\sqrt{\gamma_m^2+2K/m}\right)\nn\\
\geq&\,D(P\|Q).\nn
\end{align}
The last inequality follows from Theorem~\ref{thm:sufficient_UHT} because  $\gamma_m^2+2K/m\to 0$ when $m\to\infty$. Applying Chernoff-Stein lemma completes the proof.
\end{IEEEproof}

It is worth noting that the tests in this section (e.g., Example~\ref{exp1}) require closed-form integrals, namely, $\mathbf E_{x,x'}k(x,x')$ and $\mathbf E_xk(x,y_i)$, which may be difficult to obtain for non-Gaussians. Our purpose here is to show that the universally optimal type-II error exponent is indeed achievable for non-finite sample spaces, providing a meaningful optimality criterion for nonparametric one-sample tests. In the next section, we consider another class of MMD based tests without the need of closed-form integrals.
	
\subsection{Kernel Two-Sample Tests for One-Sample Testing}
\label{sec:twoforone}
In the context of model criticism, \cite{Lloyd2015ModelCrit} casts one-sample testing into a two-sample problem where one draws sample $x^n$ from distribution $P$. A question that arises is the choice of number of samples, which is not obvious due to the lack of an explicit criterion. In light of UHT, we may ask how many samples would suffice for a kernel two-sample test to attain the type-II error exponent $D(P\|Q)$.

Denote by $\hat P_n$ the empirical measure of $x^n$. Consider a two-sample test with acceptance region \[\Omega_0(n,m)=\{(x^n,y^m):d_k(\hat P_n,\hat Q_m)\leq\gamma_{n,m}\},\]
where $d_k(\hat P_n,\hat Q_m)$ is given in (\ref{eqn:twosampleMMD}),  $K$ is a finite bound on $k(\cdot,\cdot)$, and
\begin{align} 
\gamma_{n,m}=2(K/n)^{1/2}+2(K/m)^{1/2}+\left(-2\log(\alpha/2)(K/m+K/n)\right)^{1/2}.
\end{align}
Notice that the type-II error probability now depends on both $P$ and $Q$, due to the use of $\hat P_n$. Although additional randomness is introduced, it does not hurt the type-II error exponent provided that $n$ is sufficiently large, as stated below.

	\begin{theorem} 
		\label{thm:2sample1}
		Assume the same conditions as in Theorem~\ref{thm:simple1}, and that $x^n$ i.i.d.~$\sim P$ and $y^m$ i.i.d.~$\sim Q$. Let $\Omega_1(n,m)=\mathcal X^{n+m}\setminus\Omega_0(n,m)$ be the rejection region. Letting $n$ be such that $n/m\to\infty$ as $m\to\infty$, we have	
\begin{align}	
&\text{a) under}~H_0: P=Q,\,\mathbf P_{x^n\sim P,y^m\sim P}\left(\Omega_1(n,m) \right)\leq\alpha,\nn\\
&\text{b) under}~H_1: P\neq Q,\,\liminf_{m\to\infty}-\frac{1}{m}\log\mathbf P_{x^n\sim P,y^m\sim Q}\left(\Omega_0(n,m)\right)=D(P\|Q).\nn
\end{align}	
\end{theorem}
\begin{IEEEproof}
That the two-sample test is level $\alpha$ can be verified by Lemma~\ref{lem:gammanm}. The rest is to show the type-II error exponent being $D(P\|Q)$. To proceed, we write the type-II error probability as \[
\mathbf P_{x^n\sim P,y^m\sim Q}\left(d_k(\hat P_n,\hat Q_m)\leq\gamma_{n,m}\right)=\beta_{n,m}^u+\beta_{n,m}^l,\]
where
\begin{align}
\beta_{n,m}^u&=\mathbf P_{x^n\sim P,y^m\sim Q}\left(d_k(\hat P_n,\hat Q_m)\leq\gamma_{n,m},d_k(P,\hat P_n)>\gamma_{n,m}'\right),\nn\\
\beta_{n,m}^l&=\mathbf P_{x^n\sim P,y^m\sim Q}\left(d_k(\hat P_n,\hat Q_m)\leq \gamma_{n,m}, d_k(P,\hat P_n)\leq\gamma_{n,m}'\right)\nn,\\
\gamma_{m,n}'&=\sqrt{2K/n}+\sqrt{2KmD(P\|Q)/n}\nn
\end{align}
It suffices to show that both $\beta^u_{n,m}$ and $\beta^l_{n,m}$ decrease at least exponentially fast at a rate of $D(P\|Q)$. We first have
\begin{align}
\label{eqn:betaupper}
\beta_{n,m}^u\leq \mathbf P_{x^n\sim P,y^m\sim Q}\left(d_k(P,\hat P_n)> \gamma_{n,m}'\right)\leq P_{x^n\sim P}\left(d_k(P,\hat P_n)> \gamma_{n,m}'\right)\leq e^{-mD(P\|Q)},
\end{align}
where the  last inequality is due to Lemma~\ref{lem:gamman}. Thus, $\beta_{n,m}^u$ vanishes at least exponentially fast with the exponent $D(P\|Q)$.

For $\beta_{n,m}^l$, we have 
\begin{align}
\beta_{n,m}^l&=\mathbf P_{x^n\sim P,y^m\sim Q}\left(d_k(\hat P_n,\hat Q_m)\leq \gamma_{n,m}, d_k(P,\hat P_n)\leq\gamma_{n,m}'\right)\nn\\
&\leq\mathbf P_{x^n\sim P, y^m\sim Q}\left(d_k(\hat P_n,\hat Q_m)+d_k(P,\hat P_n)\leq\gamma_{n,m}+\gamma_{n,m}'\right)\nn\\
&\stackrel{(a)}{\leq}\mathbf P_{x^n\sim P, y^m\sim Q}\left(d_k(P,\hat Q_m)\leq\gamma_{n,m}+\gamma_{n,m}'\right)\nn\\
&\leq\mathbf P_{y^m\sim Q}\left(d_k(P,\hat Q_m)\leq\gamma_{n,m}+\gamma_{n,m}'\right), \nn
\end{align}
where $(a)$ is from the triangle inequality for metric $d_k$. Similar to (\ref{eqn:common1}), we get
\begin{align}
\liminf_{n\to\infty}-\frac1n\log\beta_{n,m}^l\geq D(P\|Q),\nn
\end{align}
because $\gamma_{n,m}+\gamma_{n,m}'\to0$ as $m\to\infty$. Together with (\ref{eqn:betaupper}), we have under $H_1:P\neq Q$,
\begin{align}
\liminf_{m\to\infty}-\frac{1}{m}\log \mathbf P_{x^n\sim P,y^m\sim Q}\left(d_k(\hat P_n,\hat Q_m)\leq\gamma_{n,m}\right)\geq D(P\|Q).\nn
\end{align}

We next show the other direction under $H_1$. We can write
\begin{align}
\mathbf P_{x^n\sim P,y^m\sim Q}\left(d_k(\hat P_n,\hat Q_m)\leq \gamma_{n,m}\right)
\stackrel{(a)}{\geq}&\,\mathbf P_{x^n\sim P,y^m\sim Q}\left(d_k(\hat P_n, P)\leq\gamma_{n}', d_k(P,\hat Q_m)\leq\gamma_{m}'\right)\nn\\
=&\,\mathbf P_{x^n\sim P}\left(d_k(\hat P_n, P)\leq\gamma_{n}'\right) \mathbf P_{y^m\sim Q}\left(d_k(P,\hat Q_m)\leq{\gamma_{m}'}\right),\nn
\end{align}
where $(a)$ is because $d_k$ is a metric, and we choose $\gamma_m'=\sqrt{2K/m}\big(1+\sqrt{\log\alpha^{-1/2}}\big)$ and $\gamma_n'=\sqrt{2K/n}\big(1+\sqrt{\log\alpha^{-1/2}}\big)$ so that $\gamma_{n,m}>\gamma_n'+\gamma_m'$. Then Lemma~\ref{lem:gamman} gives $\mathbf P_{x^n\sim P}(d_k(P,\hat P_n)\leq\gamma_n')>1-\sqrt{\alpha}$ and $\mathbf P_{y^m\sim Q}(d_k(P,\hat Q_m)\leq\gamma_m')>1-\sqrt{\alpha}$, where the latter implies that $d_k(P,\hat Q_m)\leq \gamma_m'$ is a level $\sqrt\alpha$ test for testing $H_0: x^n\sim P$ and $H_1:x^n\sim Q$ with $P\neq Q$.  Together with Chernoff-Stein Lemma, we get
\begin{align}
&\,\liminf_{m\to\infty}-\frac1m\log \mathbf P_{x^n\sim P,y^m\sim Q}\left(d_k(\hat P_n,\hat Q_m)\leq\gamma_{n,m}\right)\nn\\
\leq&\,\liminf_{m\to\infty}-\frac1m\log \left(\mathbf P_{x^n\sim P}\left(d_k(\hat P_n, P)\leq\gamma_{n}'\right) \mathbf P_{y^m\sim Q}\left(d_k(P,\hat Q_m)\leq{\gamma_{m}'}\right)\right)\nn\\
\leq&\,\liminf_{m\to\infty}-\frac1m\log\left(1-{\sqrt\alpha}\right)+\liminf_{m\to\infty}-\frac1m\log\mathbf P_{y^m\sim Q}\left(d_k(P,\hat Q_m)\leq{\gamma_{m}'}\right)\nn\\
\leq&\,D(P\|Q).\nn
\end{align}
The proof is complete.
\end{IEEEproof}
{
\begin{example}
\label{exp2}
Consider the same one-sample problem as in Example~\ref{exp1} with $P\sim\mathcal N(0,1)$. We may pick the same Gaussian kernel $k(x,y)=\exp(-(x-y)^2/2)$. Then the threshold becomes $\gamma_{n,m}=2(1/n)^{1/2}+2(1/m)^{1/2}+\left(-2\log(\alpha/2)(1/m+1/n)\right)^{1/2}$. For standard Gaussian distribution, we can easily draw $n$ i.i.d.~samples from $P$, denoted as $\{x_i\}_{i=1}^n$. The test statistic can be calculated using (\ref{eqn:twosampleMMD}) by plugging in the Gaussian kernel function $\exp(-(x-y)^2/2)$, which has $\mathcal O((n+m)^2)$ time complexity. Here we comment that i.i.d.~samples from $P$ may not be easily obtained when $P$ is complicated or is subject to a normalization constant. Fortunately, such cases may be handled by another kernel based test, as will be discussed in the next section. 
\end{example}
}

Replacing the first two terms in $d_k^2(\hat P_n,\hat Q_m)$ with $\frac{1}{m(m-1)}\sum_{i=1}^m\sum_{j\neq i} k(y_i,y_j)$ and $\frac{1}{n(n-1)}\sum_{i=1}^n\sum_{j\neq i} k(x_i,x_j)$ also results in an unbiased statistic, which we denote as $d_u^2(\hat P_n, \hat Q_m)$ \cite{Gretton2012}. We then have the following universally optimal test based on $d_u^2(\hat P_n, \hat Q_m)$.
	\begin{corollary}
		\label{cor:2sample2}
		Under the same assumptions as in Theorem~\ref{thm:2sample1}, the test defined by $\Omega_0=\{y^m:d_u^2(\hat P_n,\hat Q_m)\leq\gamma_{n,m}^2+K/n+K/m\}$ has its type-I error probability below $\alpha$ and type-II error exponent being $D(P\|Q)$, provided that ${n}/{m}\to\infty$ as $m\to\infty$.
	\end{corollary}
\begin{IEEEproof}[Proof (Sketch)]
Similar to the proof of Corollary~\ref{cor:simple2} by noting that $|d_u^2(\hat P_n,\hat Q_m)-d_k^2(\hat P_n,\hat Q_m)|\leq K/n+K/m$.
\end{IEEEproof}
\begin{remark}
The above result can be treated as a special case of the two-sample problem where samples sizes scale in different orders. For the  case with $0<\lim_{m\to\infty}n/m<\infty$, however, the current approach is not readily applicable. A naive way is to attempt to decompose the acceptance region $\Omega_0(n,m)$ into $\Omega_0'(n)\times\Omega_0''(m)$ with $\Omega_0'(n)$ and $\Omega_0''(m)$ being respectively decided by $x^n$ and $y^m$, and then apply Sanov's theorem to each set. Unfortunately, such a decomposition is not feasible for the MMD based two-sample tests. We postpone a further investigation until Section~\ref{sec:two}, after studying the KSD based one-sample tests.
\end{remark}

\subsection{Kernel Stein Discrepancy based One-Sample Tests}
\label{sec:KSD1sampletest}
As mentioned in Section~\ref{sec:preliminary}, there does not exist a uniform or distribution-free probabilistic bound on $d_S^2(P,\hat Q_m)$, and it becomes difficult to find a threshold to meet the constant level constraint for all sample sizes. To proceed, we relax the level constraint to an asymptotic one and use the result of Lemma~\ref{lem:KSDnull} which states that $md_S^2(P,\hat Q_m)$ converges weakly to some distribution under $H_0:P=Q$. We assume a fixed $\alpha$-quantile $\gamma_{\alpha}$ of the limiting cumulative distribution function, i.e., $\lim_{m\to\infty}P(md_S^2(P,\hat Q_m)>\gamma_\alpha)=\alpha$. If $\gamma_m$ is such that $\gamma_m\to0$ and $\lim_{m\to\infty}m\gamma_m\to\infty$, e.g., $\gamma_m=\sqrt{1/m}\left(1+\sqrt{-\log\alpha}\right)$, we get $m\gamma_m>\gamma_\alpha$ in the limit and thus $\lim_{m\to\infty}P(d_S^2(P,\hat Q_m)>\gamma_m)\leq\alpha$. Together with the weak convergence properties of the KSD, we have the following result. 
\begin{theorem}
	\label{thm:KSDmain}
	Let $P$ and $Q$ be distributions defined on $\mathbb R^d$, with $D(P\|Q)<\infty$ under the alternative hypothesis. Assume $y^m~\text{i.i.d.}\sim Q$ and set $\gamma_m=\sqrt{1/m}\left(1+\sqrt{-\log\alpha}\right)$. Then we have
	\begin{enumerate}
		\item assuming the conditions in Lemma~\ref{lem:KSDnull},   we have
		\[\lim_{n\to\infty} \mathbf P_{y^m\sim P}\left(d_S^2(P,\hat Q_m)> \gamma_m\right)\leq\alpha,~\text{under}~H_0;\] 	
		\item 
		assuming that kernels satisfy the conditions in Theorem~\ref{thm:KSD_weak},    then
		\[\liminf_{m\to\infty}-\frac1m\log \mathbf P_{y^m\sim Q}\left(d_S^2(P,\hat Q_m)\leq \gamma_m\right)=D(P\|Q),~\text{under}~H_1.\]
	\end{enumerate}
\end{theorem}
\begin{IEEEproof}
To establish the type-II error exponent, let $d_{W}$ denote the weak metrizable metric that lower bounds the KSD in Theorem~\ref{thm:KSD_weak}. Then  $d_{W}(P,\hat Q_m)\leq g(d_{S}(P,\hat Q_m))$ where $g(d_{S})\to 0$ as $d_{S}\to 0$. Then there exists $\gamma_m'$ such that $\{y^m:d^2_{S}(P,\hat Q_m)\leq\gamma_m\}\subset\{y^m:d^2_{W}(P,\hat Q_m)\leq\gamma_m'\}$ and $\gamma_m'\to 0$ as $m\to\infty$. The rest follows from the sufficient condition in Theorem~\ref{thm:sufficient_UHT}.
\end{IEEEproof}
{
\begin{example}
\label{exp3}
We continue with the same setting as in Examples~\ref{exp1} and \ref{exp2}, and pick the same Gaussian kernel $k(x,y)=\exp(-(x-y)^2/2)$. We then have  
\begin{align}
    s_p(x)&=\nabla_x \log p(x)=\nabla_x\log\left(\frac{1}{\sqrt{2\pi}}e^{-x^2/2}\right)=-x,\nn\\
    \nabla_x k(x,x')&=\nabla_x e^{-(x-x')^2/2} = -(x-x')e^{-(x-x')^2/2},\nn\\  \operatorname{trace}(\nabla_{x,x'} k(x,x'))& = \frac{\partial e^{-(x-x')^2/2}}{\partial x\partial x'}=((x-x')^2-1)e^{-(x-x')^2/2}. \nn
\end{align} According to (\ref{eqn:KSD_h}), the test statistic is
\begin{align}
    d_S^2(P,\hat Q_m) = \sum_{i=1}^m\sum_{j=1}^m h_p(y_i,y_j) = \sum_{i=1}^m\sum_{j=1}^m \left(2(y_i-y_j)^2+y_iy_j-1)\right) e^{-(y_i-y_j)^2/2}.\nn
\end{align}
\end{example}
Compared with  Example~\ref{exp2}, we see that there is no need to draw i.i.d.~samples of $P$. Hence, the KSD based test may be computationally convenient when the density function $p$ is given in a complicated form and/or subject to a normalization factor. Such a case is frequently encountered in Bayesian inference involving complicated posterior distributions \cite{Bishop2006PRML}.   
}

An unbiased U-statistic $d_{S(u)}^2(P,\hat Q_m)=\frac{1}{m(m-1)}\sum_{i=1}^m\sum_{j\neq i} h_p(y_i,y_j)$ for estimating $d_S^2(P,Q)$ was proposed in \cite{Liu2016GoodnessFit}. A similar result holds under an additional assumption on the boundedness of $h_p$ according to the same argument of Corollary~\ref{cor:simple2}; detailed proof is omitted.
\begin{corollary}
	Assume the same conditions as in Theorem~\ref{thm:KSDmain} and further that $h_p(\cdot,\cdot)\leq H_p$ for some $H_p\in\mathbb R^+$. Then the test $d_{S(u)}^2(P,\hat Q_m)\leq \gamma_m+H_p/m$ is asymptotically level $\alpha$ and achieves the optimal type-II error exponent $D(P\|Q)$.
\end{corollary}
\subsection{Remarks}
\label{sec:remark}
We have the following remarks regarding our results.
\subsubsection{Threshold choice} The distribution-free thresholds used in the MMD based tests are generally too conservative, as the actual distribution $P$ is not taken into account. Alternatively, one may use Monte Carlo or bootstrap methods to empirically estimate the acceptance threshold \cite{Chwialkowski2016Goodness,Gretton2012,Jitkrittum2017linearGoodness}, making the tests asymptotically level $\alpha$, i.e., $\lim_{m\to\infty}\alpha_m\leq\alpha$. Bootstrap thresholds have also been proposed for the KSD based tests in \cite{Arcones1992bootstrap,Chwialkowski2014wild,Leucht2012degenerate}.  These methods, however, introduce additional randomness on the threshold choice and further on the type-II error probability. As a result, it becomes difficult to establish the optimal type-II error exponent. A simple fix is to take the minimum of the Monte Carlo or bootstrap threshold and the distribution-free one, guaranteeing a deterministically vanishing threshold and hence the optimal type-II error exponent.  
\subsubsection{Weak metrizable property} To apply Sanov's theorem as in our approch, we find a superset of probability measures for the equivalent acceptance region, which is required to be closed and to converge (in terms of weak convergence) to $P$ in the large sample limit. Without the weak convergence property, the equivalent acceptance region may contain probability measures that are not close to $P$, and the minimum KLD over the superset would be hard to obtain. An example can be found in \cite[Theorem~6]{Gorham2017measuringKSD}  where the KSDs are driven to zero by sequences of probability measures not converging to $P$. Consequently, our approach does not establish the optimal type-II error exponent with the linear-time KSD based tests in \cite{Jitkrittum2017linearGoodness, Liu2016GoodnessFit}, the linear-time kernel two-sample test in \cite{Gretton2012}, the kernel based B-test in \cite{Zaremba2013Btest}, and a pseudometric based two-sample test in \cite{Chwialkowski2015fast}, due to the lack of the weak metrizable property.
\subsubsection{ Non-i.i.d.~data} We notice that \cite{Chwialkowski2016Goodness} considered non-i.i.d.~data by use of wild bootstrap. In general, however, statistical optimality with non-i.i.d.~data is difficult to establish even for simple hypothesis testing.

\subsection{Other Asymptotic Criteria}
Before ending this section, we would like to discuss two other related asymptotic statistical criteria.

\subsubsection{Exact Bahadur slope} We consider the exact Bahadur slope for its close relationship with our asymptotic statistical criterion \cite{Bahadur1960,Serfling}. In particular, the exact Bahadur slope for a hypothesis test is equivalent to twice of the type-I error exponent, subject to a constant constraint on the type-II error probability, that is, 
\[\liminf_{m\to\infty} -\frac{2}{m}\log\alpha_m,~\text{subject to}~\beta_m\leq\beta.\]
The optimal exact Bahadur slople is given by $2D(Q\|P)$, assuming that $0<D(Q\|P)<\infty$. However, the universal optimality w.r.t.~this criterion cannot be achieved for any one-sample test. To see this, notice that a nonparametric one-sample test, including both the test statistic and threshold, is constructed only through the sample $y^m$ and the target distribution $P$. Moreover, the type-I error probability is defined w.r.t.~the null hypothesis when $y^m\sim P$. Therefore, the type-I error exponent of a nonparametric one-sample test is characterized by only $P$ and cannot capture the information of the alternative distribution $Q$, thereby cannot attain the optimum $D(Q\|P)$ in the universal setting.
\subsubsection{Chernoff index}
The Chernoff index of a hypothesis test is defined as the minimum of its type-I and type-II error exponents, i.e., 
\begin{align}
\label{eqn:Chernoff_index}
\min \left\{\liminf_{m\to\infty} -\frac{1}{m}\log\alpha_m,\,\liminf_{m\to\infty}-\frac{1}{m}\log\beta_m\right\}.
\end{align}
Assuming that $P$ and $Q$ are mutually absolutely continuous, then the maximum Chernoff index is given by the Chernoff information and is achieved by the likelihood ratio test whose type-I and type-II error exponents are equal \cite{Cover2006,Dembo2009}. As discussed above with the exact Bahadur slope, the type-I error probability of a nonparametric test is independent of the alternative distribution $Q$, whereas the optimal Chernoff index, Chernoff information, depends on both $P$ and $Q$. Therefore, the optimal Chernoff index can not be achieved in the universal setting, either. 

\section{Asymptotically Optimal Two-Sample Tests}  
\label{sec:two}
In this section, we present our main results on the type-II error exponent for general kernel two-sample tests. As discussed in Section~\ref{sec:twoforone}, the first and the most important step is to establish an extended Sanov's theorem that works with two sample sequences. 
\subsection{Extended Sanov's Theorem}
\label{sec:two_Sanov}
We begin with our definition of  pairwise weak convergence for probability measures: we say $(P_l,Q_l)\to(P,Q)$ weakly if and only if both $P_l\to P$ and $Q_l\to Q$ weakly. Consider $\mathcal P\times\mathcal P$ endowed with the topology induced by this pairwise weak convergence; it can be verified that this topology is equivalent to the product topology on $\mathcal P\times\mathcal P$ where each $\mathcal P$ is endowed with the topology of weak convergence. An extended version of Sanov's theorem handling two distributions is stated below, which may be of independent interest to other large deviation applications.

\begin{theorem}[Extended Sanov's Theorem] 
	Let $\mathcal X$ be a Polish space, $x^n$~i.i.d.~$\sim P$, and $y^m$~i.i.d.~$\sim Q$. Assume $0<\lim_{n,m\to\infty}\frac{n}{n+m}=:c<1$. Then for a set $\Gamma\subset\mathcal P\times\mathcal P$, 
	\begin{align}
\limsup_{n,m\to\infty}-\frac{1}{n+m}\log\mathbf {P}_{x^n\sim P, y^m\sim Q}((\hat{P}_n,\hat{Q}_m)\in\Gamma)&\leq 	\inf_{(R,S)\in\operatorname{int}\Gamma} cD(R\|P)+(1-c)D(S\|Q),\nn\\
	\liminf_{n,m\to\infty}-\frac{1}{n+m}\log\mathbf P_{x^n\sim P,y^m\sim Q} ((\hat{P}_n,\hat{Q}_m)\in\Gamma)&\geq\inf_{(R,S)\in\operatorname{cl}\Gamma} cD(R\|P)+(1-c)D(S\|Q),\nn
	\end{align}
	where $\operatorname{int}\Gamma$ and $\operatorname{cl}\Gamma$ are the interior and closure of $\Gamma$ w.r.t.~the pairwise weak convergence, respectively.
\end{theorem}
\begin{IEEEproof}
See Appendix~\ref{sec:extendedSanov}.  
\end{IEEEproof}	

We comment that the extended Sanov's theorem is not apparent from the original one, as existing tools, e.g., Cram{\' e}r theorem \cite{Dembo2009} that is used for proving the original Sanov's theorem, can only handle a single distribution. {Nevertheless, inspired by \cite{Csiszar2006simple} that proved Sanov's theorem (w.r.t.~a single distribution) in the $\tau$-topology, we first establish the above result in finite sample spaces and then extend it to general Polish spaces, with two simple combinatorial lemmas as prerequisites.}

\subsection{Exact and Optimal Error Exponent}
\label{sec:ExpConsistency}
With the extended Sanov's theorem and a vanishing threshold, we are ready to establish the type-II error exponent for the kernel two-sample test defined in Section~\ref{sec:twoforone}. Our result follows.
\begin{theorem}
	\label{thm:mainresult1}
	Assume \hyperref[thm:MMDmetrize]{\bf A1}, \hyperref[thm:MMDmetrize]{\bf A2}, and  $0<\lim_{n,m\to\infty}\frac{n}{n+m}=:c<1$. Further assume that \[0<D^*\coloneqq \inf_{R\in\mathcal P} cD(R\|P) + (1-c)D(R\|Q)<\infty,\] under the alternative hypothesis $H_1$. Then the kernel test
	$d_k(\hat P_n, \hat Q_m)\leq \gamma_{n,m}$,	with $d_k(\hat P_n, \hat Q_m)$ and $\gamma_{n,m}$ are defined in Section~\ref{sec:twoforone}, is an exponentially consistent level $\alpha$ test with type-II error exponent being $D^*$, that is, 
	\[\alpha_{n,m}\leq\alpha~\text{and}~\liminf_{n,m\to\infty}-\frac1{n+m}\log\beta_{n,m}=D^*.\]
\end{theorem}
\begin{IEEEproof}
A proof is given in Appendix~\ref{sec:proofappendix}, which is similar to the proof of Theorem~\ref{thm:sufficient_UHT} but with some extra treatment on the pairwise weak convergence .
\end{IEEEproof}

Therefore, when $0<c<1$, the type-II error probability  vanishes as $\mathcal O(e^{-(n+m)(D^*-\epsilon)})$, where $\epsilon\in(0,D^*)$ is fixed and can be arbitrarily small. The result also shows that the choice of kernels only affects the sub-exponential term in the type-II error probability, provided that they meet the conditions of \hyperref[thm:MMDmetrize]{\bf A2}.

Now that we have obtained the exact type-II error exponent for the kernel two-sample test, we proceed to derive an upper bound on the optimal type-II error exponent for any (asymptotically) level $\alpha$ test.
\begin{theorem}
	\label{thm:upperbd}
	Let $x^n$, $y^m$, $P$, $Q$, and $D^*$ be defined as in Theorem~\ref{thm:mainresult1}.  If a nonparametric two-sample test $\Omega'(n,m)$  is (asymptotically) level $\alpha,0<\alpha<1$,  its type-II error probability $\beta'_{n,m}$ satisfies:
	\begin{itemize}
	\item if~$0<\lim_{n,m\to\infty}\frac{n}{n+m}=c<1$ and $0<D^*<\infty$, then 
\[\liminf_{n,m\to\infty}-\frac{1}{n+m}\log\beta'_{n,m}\leq D^*;\]
\item if $\lim_{n,m\to\infty}\frac nm=\infty$ and $0<D(P\|Q)<\infty$, then
	\[\liminf_{n,m\to\infty}-\frac{1}{m}\log\beta'_{n,m}\leq D(P\|Q).\]
	\end{itemize}
\end{theorem}	
\begin{IEEEproof}
Our proof is based on the notion of relative entropy typical set, where the relative entropy is another name for the KLD \cite{Cover2006}.

We begin with the case where $0<c<1$. Let $P'$ be such that $cD(P'\|P)+(1-c)D(P'\|Q)=D^*$. Consider first $D(P'\|P)\neq0$ and $D(P'\|Q)\neq0$. Since $D^*$ is assumed to be finite, we have both $D(P'\|P)$ and $D(P'\|Q)$ being finite. This implies that $P'$ is absolutely continuous w.r.t.~both $P$ and $Q$, so the Radon-Nikodym derivatives ${dP'}/{dP}$ and ${dP'}/{dQ}$ exist.  

We can define the following set
	\begin{equation}
	\begin{aligned}
	\label{eqn:kldset}
	A_n = \left\{x^n:D(P'\|P)-\epsilon\leq\frac{1}{n}\log\frac{dP'(x^n)}{dP(x^n)}\leq D(P'\|P)+\epsilon\right\},
	\end{aligned}
	\end{equation}
which contains samples so that the log-likelihood ratios are $\epsilon$-close to the true KLD, and is called the relative entropy typical set. Recall the definition of the KLD: $D(P'\|P)=\mathbf{E}_{x\sim P'}\log(dP'(x)/dP(x))$. By law of large numbers,  we have for any given $\epsilon>0$, 
\[\mathbf P_{x^n\sim P'}(A_n)\geq1-\epsilon/2,~\text{for large enough}~n.\]
Similarly, define 
	\begin{equation}
\begin{aligned}
\label{eqn:kldset2}
B_m = \left\{y^m:D(P'\|Q)-\epsilon\leq\frac{1}{m}\log\frac{dP'(y^m)}{dQ(y^m)}\leq D(P'\|Q)+\epsilon\right\},
\end{aligned}
\end{equation}
and we have
\[\mathbf P_{y^m\sim P'}(B_m)\geq1-\epsilon/2,~\text{for large enough}~m.\]
Therefore, we get
	\begin{align}
	\label{eqn:kldsetlargeProb}
	\mathbf P_{x^n\sim P',y^m\sim P'}(A_n\times B_m)=\mathbf P_{x^n\sim P',y^m\sim P'}(x^n\in A_n, y^m\in B_m)\geq 1-\epsilon,~\text{for large enough}~n~\text{and}~m.
	\end{align}
	
Now consider the type-II error probability for a level $\alpha$ test. First, if a test is level $\alpha$, we have its acceptance region satisfy
	\begin{align}
	\label{eqn:anyalphatestLargeProb}
	\mathbf P_{x^n\sim P, y^m\sim P}\left(\Omega_0'(n,m)\right)> 1-\alpha,
	\end{align}
when the null hypothesis $H_0$ holds, i.e., when $x^n$ and $y^m$ are i.i.d. according to a common distribution (which is not necessarily $P'$). Then under the alternative hypothesis $H_1:P\neq Q$, we have 
	\begin{align}
	\beta'_{n,m}&= \mathbf P_{x^n\sim P, y^m\sim Q} \left(\Omega'_0(n,m)\right)\nn\\
	&\geq\mathbf P_{x^n\sim P, y^m\sim Q} \left (A_n\times B_m\cap\Omega'_0(n,m)\right)\nn\\
	&= \int_{A_n\times B_m\cap\Omega'_0(n,m)} dP(x^n)\,dQ(y^m)\nn\\
	&\stackrel{(a)}{\geq}\int_{A_n\times B_m\cap\Omega'_0(n,m)}2^{-n(D(P'\|P)+\epsilon)}2^{-m(D(P'\|Q)+\epsilon)} dP'(x^n)\,dP'(y^m)\nn\\
	&=2^{-nD(P'\|P)-m(D(P'\|Q)-(n+m)\epsilon} \int_{A_n\times B_m\cap \Omega'_0(n,m)}dP'(x^n)\,dP'(y^m)\nn\\
	&\stackrel{(b)}{\geq} 2^{-nD(P'\|P) -mD(P'\|Q)-(n+m)\epsilon}(1-\alpha-\epsilon),\nn
	\end{align}
	where $(a)$ is from the defintion of $A_n$ and $B_m$, and $(b)$ is due to (\ref{eqn:kldsetlargeProb}) and (\ref{eqn:anyalphatestLargeProb}). Thus, when $\epsilon$ is small enough so that $1-\alpha-\epsilon>0$, we get 
	\begin{align}
	\label{eqn:upp}
	\liminf_{n,m\to\infty}-\frac{1}{n+m}\log\beta'_{n,m}&\leq \liminf_{n,m\to\infty}\frac{1}{n+m}\left(nD(P'\|P)+m(D(P'\|Q)+(n+m)\epsilon\right)\nn\\&=D^*+\epsilon.
	\end{align}
	
	If a test is an asymptotic level $\alpha$ test, we can replace $\alpha$ by $\alpha+\epsilon'$ where $\epsilon'$ can be made arbitrarily small provided that $n$ and $m$ are large enough. Thus, the above equation (\ref{eqn:upp}) holds, too. Since $\epsilon$ can also be arbitrarily small, we conclude that \[\liminf_{n,m\to\infty}-\frac{1}{n+m}\log\beta'_{n,m}\leq D^*.\]		
	
If $P'=P$, then $A_n$ contains all $x^n\in\mathcal X^n$ and the above procedure results in the same result. 
	
Finally, the same argument also applies the case with $\lim_{n,m\to\infty}\frac{n}{m}=\infty$ and we have
	\[\liminf_{n,m\to\infty}-\frac{1}{m}\log\beta'_{n,m}\leq D(P\|Q).\]
\end{IEEEproof}

This theorem shows that the kernel test $d_k(\hat P_n,\hat Q_m)\leq \gamma_{n,m}$ is an optimal level $\alpha$ two-sample test, by choosing the type-II error exponent as the asymptotic performance metric. Moreover, Theorems~\ref{thm:mainresult1} and \ref{thm:upperbd} together provide a way of identifying more universally optimal two-sample tests:
\begin{itemize}
	\item Assuming $n=m$, the  test $d_u^2(\hat P_n, \hat Q_m)\leq(4K/\sqrt{n})\sqrt{\log(\alpha^{-1})}$ is also level $\alpha$ \cite[Corollary~11]{Gretton2012}. As $k(\cdot,\cdot)$ is finitely bounded by $K$, its type-II error probability vanishes exponentially at a rate of $\inf_{R\in\mathcal P}\frac{1}{2}D(R\|P)+\frac12D(R\|Q)$, which can be shown by the same argument of Corollary~\ref{cor:simple2}.
	
	
	\item It is also possible to consider a family of kernels for the test statistic \cite{Fukumizu2009,Sriperumbudur2016EstPM}. For a given family $\kappa$, the test statistic is $\sup_{k\in\kappa} d_k(\hat P_n, \hat Q_m)$ which also metrizes weak convergence under suitable conditions, e.g., when $\kappa$ consists of finitely many Gaussian kernels \cite[Theorem~3.2]{Sriperumbudur2016EstPM}. If $K$ remains to be an upper bound for all $k\in\kappa$, then comparing $\sup_{k\in\kappa} d_k(\hat P_n, \hat Q_m)$ with $\gamma_{n,m}$ defined in Section~\ref{sec:twoforone} results in an asymptotically optimal level $\alpha$ test.
\end{itemize}
\begin{remark}
There is a similar result to Lemma~\ref{lem:gamman}  for the unbiased two-sample statistic $d_u^2(\hat P_n, \hat Q_m)$. Assume that $0 < \lim_{n,m\to\infty} n/(n+m)<1$ and kernel $k(\cdot,\cdot)$ is bounded by $K$, then \cite[Theorem~12]{Gretton2012OptKernelLarge} shows that the statistic $(n+m)d_u^2(\hat P_n, \hat Q_m)$ converges in distribution to some distribution under the null hypothesis. With a fixed $\alpha$-quantile $\gamma_{\alpha}'$ for the limiting distribution, then $(m+n)d_u^2(\hat P_n, \hat Q_m)\leq\gamma_\alpha'$ is level $\alpha$ in the sample limit. Consequently, the (asymptotic) level $\alpha$ constraint requires the threshold for $d_u^2(\hat P_n, \hat Q_m)$  decrease at most $\mathcal O(1/(n+m))$ fast. 
\end{remark}
\begin{remark} In \cite{Ramdas2015}, a notion of fair alternative is proposed for two-sample testing as dimension increases, which is to fix $D(P\|Q)$ under the alternative hypothesis for all dimensions. This idea is guided by the fact that the KLD is a fundamental information-theoretic quantity  determining the hardness of hypothesis testing problems. This approach, however, does not take into account the impact of sample sizes. In light of our results, perhaps a better choice is to fix $D^*$ in Theorem~\ref{thm:mainresult1} when the sample sizes grow in the same order. In practice, $D^*$ may be hard to compute, so fixing its upper bound $(1-c)D(P\|Q)$ and hence $D(P\|Q)$ is reasonable.
\end{remark}
\begin{remark}
	\label{rmk:kernelchoice}
The main results indicate that the type-II error exponent is independent of the choice of kernels as long as kernels are bounded continuous and characteristic. We remark that this indication does not contradict previous studies on kernel choice, as the sub-exponential term can dominate in the finite sample regime. In light of the exponential decay, it then raises interesting connections with a kernel selection strategy, where part of samples are used as training data to choose a kernel and the remaining samples are used with the selected kernel to compute the test statistic \cite{Gretton2012OptKernelLarge,Sutherland2017GeneandCrit}. On the one hand, the sample size should not be too small so that there are enough data for training. On the other hand, if the number of samples is large enough and the exponential decay term becomes dominating, directly using the entire samples may be good enough to have a low type-II error probability, provided that kernel is not too poor. We conduct a toy experiment to further illustrate this point in Appendix~\ref{sec:exp}. Selecting a proper kernel is an important ongoing research topic and we  refer the reader to existing works on kernel selection strategy, e.g., \cite{Gretton2012OptKernelLarge,Sutherland2017GeneandCrit}.
\end{remark}

\section{Application to Off-line Change Detection}
\label{sec:app}
In this section, we apply our results to off-line change detection.  To our best knowledge,  no tests have been shown to be optimal, in terms of either error probability or error exponent, for detecting the change when no prior information on the post-change distribution  is available \cite{Basseville1993detectionchanges}. We only study the case where both the pre- and post-change distributions are unknown; the case with a known pre-change distribution can be handled similarly and is omitted.

Let $z_1,\ldots,z_{n}\in\mathbb{R}^d$ be an independent sequence of observations. Assume that there is at most one change-point at index $1<t<n$, which, if exists, indicates that $z_i\sim P, 1\leq i\leq t$ and $z_i\sim Q, t+1\leq i\leq n$ with $P\neq Q$. The off-line change-point analysis consists of two steps: 1) detect if there is a change-point in the sample sequence; 2) estimate the index $t$ if such a change-point exists. Notice that a method may readily extend to multiple change-point and on-line settings, through sliding windows running along the sequence, as in \cite{Desobry2005onlinekernel, Harchaoui2009KernelChange,Li2015Changedetection}.

The first step in the change-point analysis is usually formulated as a hypothesis testing problem: 
\begin{align}
H_0:~&z_i \sim P, i=1,\ldots,n, \nn\\
H_1:~&\text{there exists}~1<t<n~\text{such that}\nn\\
&z_i\sim P, 1\leq i\leq t~\text{and}~z_i\sim Q\neq P, t+1\leq i\leq n.\nn
\end{align}
Let $\hat P_{i} $ and $\hat Q_{n-i}$ denote the empirical measures of sequences $z_1,\ldots, z_i$ and $z_{i+1},\ldots,z_n$, respectively. Then an MMD based test can be directly constructed using the  maximum partition strategy:
\[\textrm{decide}~H_0,~\text{if}~\max_{a_n\leq i\leq b_n}d_k(\hat P_{i}, \hat Q_{n-i})\leq\gamma_n,\]
where the maximum is searched in the interval $[a_n, b_n]$ with $a_n > 1$ and $b_n < n$. If the test favors $H_1$, we can proceed to estimate the change-point index by $\argmax_{a_n\leq i\leq b_n} d_k(\hat P_i, \hat Q_{n-i})$. Here we characterize the performance of detecting the presence of a change for this test, using Theorems~\ref{thm:mainresult1} and \ref{thm:upperbd}. We remark that the assumptions on the search interval and on the change-point index in the following theorem are standard practice for nonparametric change detection \cite{Basseville1993detectionchanges,Desobry2005onlinekernel, Harchaoui2009KernelChange,James1987testsforchange,Li2015Changedetection}.

\begin{theorem}
	\label{thm:changedetection}
	Let $0<u <v <1$, $a_n/n\to u$ and $b_n/n\to v$ as $n\to\infty$. Under the alternative hypothesis $H_1$, assume that the change-point index $t$ satisfies $u<\lim_{n\to\infty}t/n=c<v$, and that $0< D^*<\infty$ where $D^*$ is defined w.r.t.~$P$, $Q$, and $c$ in Theorem~\ref{thm:mainresult1}. Further assume that the kernel $k$ satisfies \hyperref[thm:MMDmetrize]{\bf A2}, with $K>0$ being an upper bound. Given $0<\alpha<1$, set $c_n=\min\{a_n(n-a_n), b_n(n-b_n)\}$ and $\gamma_n=2\sqrt{{K}/{a_n}}+2\sqrt{K/(n-b_n)}+\sqrt{2Kn\log(2n\alpha^{-1})/c_n}$. Then the test $\max_{a_n\leq i\leq b_n}d_k(\hat P_{i}, \hat Q_{n-i})\leq\gamma_n$ is level $\alpha$ and achieves the optimal type-II error exponent. That is,
	\[\alpha_{n}\leq \alpha,~\text{and}~\liminf_{n\to\infty}-\frac{1}{n}\log \beta_n=D^*,\]
	where $\alpha_{n}$ and $\beta_n$ are the type-I and type-II error probabilities, respectively.
\end{theorem}
\begin{IEEEproof}
We first have 
\[\mathbf P_{z^n\sim P}\left(\max_{a_n\leq i\leq b_n}d_k(\hat P_{i}, \hat Q_{n-i})>\gamma_n\right)\leq \sum_{a_n\leq i\leq b_n}\mathbf P_{z^n\sim P}\left(d_k(\hat P_{i}, \hat Q_{n-i})>\gamma_n\right).\]
To meet the type-I error constraint, it suffices to make each $\mathbf P_{z^n}(d_k(\hat P_{i}, \hat Q_{n-i})>\gamma_n)\leq \alpha/n$ under the null hypothesis $H_0$. This can be verified using Lemma~\ref{lem:gammanm} with the choice of $\gamma_n$  in the above theorem. To see the optimal type-II error exponent, consider a simpler problem where the possible change-point $t$ is known, i.e., a two-sample problem between $z_1,\ldots,z_t$ and $z_{t+1},\ldots,z_n$. Since $\gamma_n\to0$ as $n\to\infty$, applying Theorems~\ref{thm:mainresult1} and \ref{thm:upperbd} establishes the optimal type-II error exponent.
\end{IEEEproof}

	\section{Conclusion and Discussion}
\label{sec:conclusion} 
In this paper, we have established the statistical optimality of the MMD and KSD based one-sample tests in the spirit of universal hypothesis testing. {The KSD based tests can be more computationally efficient when the the density function is given in complicated forms and/or subject to a normalization constant, as there is no need to draw samples or compute integrals.} In contrast, the MMD based tests are more statistically favorable, as they require weaker assumptions and can meet the level constraint for any sample size. Following the same optimality criterion, we further show that the quadratic-time MMD based two-sample tests are also asymptotically optimal in the universal setting, by extending the Sanov's theorem to the two-sample case. Our results provide a practically meaningful  approach for constructing universally optimal one- and two-sample tests. 

A future direction is to generalize the result to a Polish sample space, without the locally compact Hausdorff assumption \cite{Zeitouni1991universal}. Although we cannot establish this result in the current work, we believe that our approach would be feasible once a proper metric is found to meet the  two conditions in Theorem~\ref{thm:sufficient_UHT}, since both Sanov's theorem and its extended version are established w.r.t.~the Polish space. 
	

\appendices
\section{Proof of the Extended Sanov's Theorem}
\label{sec:extendedSanov}
We prove the extended Sanov's theorem with a finite sample space and then extend the result to a general Polish space. Our proof is inspired by \cite{Csiszar2006simple} which proved Sanov's theorem (w.r.t.~a single distribution) in the $\tau$-topology. The prerequisites are two combinatorial lemmas that are standard tools in information theory.

For a positive integer $t$, let $\mathcal P_n(t)$ denote the set of probability distributions defined on $\{1,\ldots, t\}$ of form $P=\left(\frac{n_1}n,\cdots,\frac{n_t}n\right)$, with integers $n_1,\ldots, n_t$. Stated below are the two lemmas.
\begin{lemma}[{\cite[Theorem~11.1.1]{Cover2006}}]
	\label{lem:numEmpDistribution} The cardinality	$|\mathcal P_n(t)|\leq(n+1)^t.$
\end{lemma}
\begin{lemma}[{\cite[Theorem~11.1.4]{Cover2006}}]
	\label{lem:typeprob}
	Assume $x^n$ i.i.d.~$\sim Q$ where $Q$ is a distribution defined on $\{1,\ldots,t\}$. For any $P\in\mathcal P_n(t)$, the probability of the empirical distribution $\hat P_n$ of $x^n$ equal to $P$ satisfies	
	\[(n+1)^{-t}e^{-nD(P\|Q)}\leq \mathbf P_{x^n\sim Q}(\hat P_n=P)\leq e^{-nD(P\|Q)}.\]
\end{lemma}
\subsection{Finite Sample Space}
\subsubsection{Upper bound} Let $t$ denote the cardinality of $\mathcal X$. Without loss of generality, assume that \[\inf_{(R,S)\in\operatorname{int}\Gamma} cD(R\|P)+(1-c) D(S\|Q)<\infty,\] which indicates that  the open set $\operatorname{int}\Gamma$ is non-empty. As $0<\lim_{n,m\to\infty}\frac{n}{n+m}=c<1$, we can find $n_0$ and $m_0$ such that there exists $(P'_n,Q'_m)\in\operatorname{int}\Gamma\cap \mathcal P_n(t)\times \mathcal P_m(t)$ for all $n>n_0$ and $m>m_0$, and that $cD(P_n'\|P)+(1-c)D(Q_m'\|Q)\to\inf_{(R,S)\in\operatorname{int}\Gamma} cD(R\|P)+(1-c) D(S\|Q)$ as $n,m\to\infty$. Then we have, with $n>n_0$ and $m>m_0$,
\begin{align}
&~~~~\mathbf{P}_{x^n\sim P, y^m\sim Q}((\hat P_n, \hat Q_m)\in\Gamma)\nn\\
&=\sum_{(R,S)\in\Gamma\,\cap\, \mathcal{P}_{n}(t)\times\mathcal P_m(t)} \mathbf{P}_{x^n\sim P, y^m\sim Q}(\hat P_n=R, \hat Q_m=S)\nn\\
&\geq\sum_{(R,S)\in\operatorname{int}\Gamma\,\cap\, \mathcal{P}_{n}(t)\times\mathcal P_m(t)} \mathbf{P}_{x^n\sim P, y^m\sim Q}(\hat P_n=R, \hat Q_m=S)\nn\\
&\geq \mathbf{P}_{x^n\sim P, y^m\sim Q}(\hat P_n=P_n', \hat{Q}_m=Q_m')\nn\\
&= \mathbf P_{x^n\sim P}(\hat P_n=P'_n)\,\mathbf P_{y^m\sim Q}(\hat Q_m=Q'_m)\nn\\
&\geq(n+1)^{-t}(m+1)^{-t}e^{-nD(P_n'\|P)}e^{-mD(Q_m'\|Q)}\nn,
\end{align}
where the last inequality is from Lemma~\ref{lem:typeprob}. Then we have \begin{align}
&~~~~\limsup_{n,m\to\infty}-\frac{1}{n+m}\log\mathbf{P}_{x^n\sim P, y^m\sim Q} ((\hat{P}_n,\hat{Q}_m)\in\Gamma)\nn\\&\leq \lim_{n,m\to\infty}\frac1{n+m}\left(-t\log((n+1)(m+1))+nD(P'_n\|P)+mD(Q'_m\|Q)\right)\nn\\
&=\lim_{n,m\to\infty}\frac1{n+m}\left(nD(P'_n\|P)+mD(Q'_m\|Q)\right)\nn\\
&=\inf_{(R,S)\in\operatorname{int}\Gamma}\left(cD(R\|P)+(1-c) D(S\|Q)\right).\nn
\end{align}

\subsubsection{Lower bound} We can write the probability as
\begin{align}
&~~~~\mathbf{P}_{x^n\sim P, y^m\sim Q}((\hat P_n, \hat Q_m)\in\Gamma) \nn\\
&= \sum_{(R,S)\in\Gamma\cap \mathcal{P}_{n}(t)\times\mathcal P_m(t)} \mathbf{P}_{x^n\sim P}(\hat P_n=R)\,\mathbf{P}_{y^m\sim Q}(\hat Q_m=S)\nn\\
&\stackrel{(a)}{\leq} \sum_{(R,S)\in\Gamma\cap \mathcal{P}_n(t)\times\mathcal{P}_m(t)} e^{-nD(R\|P)}e^{-mD(S\|Q)}\nn\\
&\stackrel{(b)}{\leq} (n+1)^{t} (m+1)^{t}\sup_{(R,S)\in\Gamma} e^{-nD(R\|P)}e^{-mD(S\|Q)},
\end{align}
where $(a)$ and $(b)$ are due to  Lemmas~\ref{lem:typeprob} and \ref{lem:numEmpDistribution}, respectively. This gives \[\liminf_{n\to\infty}-\frac{1}{n+m}\log \mathbf{P}_{x^n\sim P, y^m\sim Q}((\hat{P}_n,\hat{Q}_m)\in\Gamma)\geq \inf_{(R,S)\in\Gamma} cD(R\|P)+(1-c)D(S\|Q),\]
and hence the lower bound by noting that $\Gamma\in\operatorname{cl}\Gamma$. Indeed, when the right hand side is finite, the infimum over $\Gamma$ equals the infimum over $\operatorname{cl}\Gamma$ as a result of the continuity of KLD for finite sample spaces.

\subsection{Polish sample space}
We consider the general case with $\mathcal X$ being a Polish space. Now $\mathcal{P}$ is the space of probability measures defined on $\mathcal X$ endowed with the topology of weak convergence. To proceed, we introduce another topology on $\mathcal P$ and an equivalent definition of the KLD.

{\it $\tau$-topology}: Denote by $\Pi$ the set of all partitions $\mathcal A=\{A_1,\ldots, A_t\}$ of $\mathcal X$ into a finite number of measurable sets $A_i$. For $P\in\mathcal P$, $\mathcal A\in\Pi$, and $\zeta>0$, denote 
\begin{align}
\label{eqn:opentautoplogy}
U(P,\mathcal A, \zeta) = \{P'\in\mathcal P:|P'(A_i)-P(A_i)|<\zeta, i=1,\dots,t\}.
\end{align}
The $\tau$-topology on $\mathcal P$ is the coarsest topology in which the mapping $P\to P(F)$ are continuous for every measurable set $F\subset\mathcal X$. A base for this topology is the collection of the sets (\ref{eqn:opentautoplogy}). We will use $\mathcal P_\tau$ when we refer to $\mathcal P$ endowed with this $\tau$-topology, and write the interior and closure of a set $\Gamma\in\mathcal P_\tau$ as $\operatorname{int}_\tau\Gamma$
and $\operatorname{cl}_\tau\Gamma$, respectively. We remark that the $\tau$-topology is stronger than the weak topology: any open set in $\mathcal P$ w.r.t.~weak topology is also open in $\mathcal P_\tau$ (see more details in \cite{Csiszar2006simple,Dembo2009}). The product topology on $\mathcal P_\tau\times\mathcal P_\tau$ is determined by the base of the form of 
\[U(P,\mathcal A_1, \zeta_1)\times U(Q, \mathcal A_2, \zeta_2),\]
for $(P,Q)\in\mathcal P_\tau\times\mathcal P_\tau$, $\mathcal A_1,\mathcal A_2\in\Pi$, and $\zeta_1,\zeta_2>0$. We still use $\operatorname{int}_{\tau}(\Gamma)$ and $\operatorname{cl}_{\tau}(\Gamma)$ to denote the interior and closure of a set $\Gamma\subset\mathcal P_\tau\times\mathcal P_\tau$. As there always exists $\mathcal A\in\Pi$ that refines both $\mathcal A_1$ and $\mathcal A_2$, any element from the base has an open subset \[\tilde{U}(P,Q,\mathcal A,\zeta)\coloneqq U(P,\mathcal A, \zeta)\times U(Q, \mathcal A, \zeta)\subset\mathcal P_\tau\times\mathcal P_\tau,\]
for some $\zeta >0$. 

{\it Another definition of the KLD}: We now introduce an equivalent definition of the KLD
\begin{align}
D(P\|Q)=\sup_{\mathcal A\in\Pi} \sum_{i=1}^t P(A_i)\log\frac{P(A_i)}{Q(A_i)}=\sup_{\mathcal A\in\Pi}D(P^{\mathcal A}\|Q^{\mathcal A}),\nn
\end{align}
with the conventions $0\log 0=0\log\frac{0}{0}=0$ and $a\log\frac{a}{0}=+\infty$ if $a>0$. Here $P^{\mathcal A}$ denotes the discrete probability measure $(P(A_1),\ldots,P(A_t))$ obtained from probability measure $P$ and partition $\mathcal A$. It is not hard to verify that for $0<c<1$,
\begin{align}
\label{eqn:KLDdef}
cD(R\|P)+(1-c)D(S\|Q)&=c\sup_{\mathcal{A}_1\in\Pi}D(R^{\mathcal A_1}\|P^{\mathcal A_1})+(1-c)\sup_{\mathcal A_2\in\Pi}D(S^{\mathcal A_2}\|Q^{\mathcal A_2})\nn\\
&=\sup_{\mathcal{A}\in\Pi}\left(cD\left(R^{\mathcal A}\|P^{\mathcal A}\right)+(1-c)D\left(S^{\mathcal A}\|Q^{\mathcal A}\right)\right),
\end{align}
due to the existence of $\mathcal{A}$ that refines both $\mathcal{A}_1$ and $\mathcal A_2$ and the log-sum inequality \cite{Cover2006}.

We are ready to show the extended Sanov's theorem with Polish space.

\subsubsection{Upper bound}
It suffices to consider only non-empty open $\Gamma$. If $\Gamma$ is open in $\mathcal P\times\mathcal P$, then $\Gamma$ is also open in $\mathcal P_\tau\times\mathcal P_\tau$. Therefore, for any $(R,S)\in\Gamma$, there exists a finite (measurable) partition $\mathcal A= \{A_1,\ldots,A_t\}$ of $\mathcal X$ and $\zeta>0$ such that 
\begin{align}
\label{eqn:opensubset}
\tilde{U}(R,S,\mathcal A,\zeta)=
\left\{(R',S'):|R(A_i)-R'(A_i)|<\zeta,|S(A_i)-S'(A_i)|<\zeta,i=1,\ldots,t\right\}\subset\Gamma.
\end{align}

Define the function $T:\mathcal X\to\{1,\ldots,t\}$ with $T(x)=i$ for $x\in A_i$. Then $(\hat P_n, \hat Q_m)\in\tilde{U}(R,S,\mathcal A,\zeta)$ with $R,S\in\Gamma$ if and only if the empirical measures $\hat P^{\circ}_n$ of $\{T(x_1),\ldots, T(x_n)\}\coloneqq T(x^n)$ and $\hat Q^{\circ}_m$ of $\{T(y_1),\ldots, T(y_m)\}\coloneqq T(y^m)$ lie in 
\[U^{\circ}(R,S,\mathcal A, \zeta)=\{(R^\circ,S^{\circ}):|R^{\circ}(i)-R(A_i)|<\zeta, |S^\circ(i)-S(A_i)|<\zeta,i=1,\ldots, t\}\subset \mathbb R^t\times\mathbb R^t.\]
Thus, we have
\begin{align}\mathbf{P}_{x^n\sim P, y^m\sim Q}((\hat P_n, \hat Q_m)\in\Gamma)&\geq\mathbf{P}_{x^n\sim P, y^m\sim Q}((\hat P_n, \hat Q_m)\in\tilde{U}(R,S,\mathcal A, \zeta))\nn\\
&=\mathbf{P}_{T(x^n)T(y^m)}((\hat P_n^{\circ}, \hat Q_m^{\circ})\in U^{\circ}(R,S,\mathcal A, \zeta)).\nn
\end{align}
As $T(x)$ and $T(y)$ takes values from a finite alphabet and $U^{\circ}(R,S,\mathcal A, \zeta)$ is open, we obtain that 
\begin{align}
&~\limsup_{n\to\infty}-\frac{1}{n+m}\log\mathbf{P}_{x^n\sim P, y^m\sim Q}((\hat P_n,\hat Q_m)\in\Gamma)\nn\\\leq&~ \limsup_{n\to\infty}-\frac{1}{n+m}\log\mathbf{P}_{T(x^n)T(y^m)}((\hat P_n^{\circ},\hat Q_m^{\circ})\in U^{\circ}(R,S,\mathcal A, \zeta))\nn\\
\leq&~\inf_{(R^\circ,S^\circ)\in U^{\circ}(R,S,\mathcal A, \zeta)}\left(cD(R^\circ\|P^{\mathcal A})+(1-c)D(S^\circ\|Q^{\mathcal A})\right)\nn\\
=&~\inf_{(R',S')\in\tilde{U}(R,S,\mathcal A, \zeta)} \left(cD(R'^{\mathcal A}\|P^{\mathcal A})+(1-c)D(S'^{\mathcal A}\|Q^{\mathcal A})\right)\nn\\
\leq&~ cD(R\|P)+(1-c)D(S\|Q),
\end{align}
where we have used definition of KLD in Eq.~(\ref{eqn:KLDdef})  and $(R,S)\in\tilde{U}(R,S,\mathcal A, \zeta)$ in the last inequality.  As $(R,S)$ is arbitrary in $\Gamma$, the lower bound is established by taking infimum over $\Gamma$.

\subsubsection{Lower bound} With notations
\[\Gamma^{\mathcal A}=\{(R^{\mathcal{A}},S^\mathcal{A}):(R,S)\in\Gamma\},~ \Gamma(\mathcal A)=\{(R,S):(R^{\mathcal  A},S^{\mathcal{A}})\in\Gamma^{\mathcal{A}}\},\]
where $\mathcal A=\{A_1,\ldots,A_t\}$ is a finite partition, we have
\begin{align}
&~\mathbf{P}_{x^n\sim P, y^m\sim Q}((\hat P_n,\hat Q_m)\in\Gamma)\nn\\
\leq&~\mathbf{P}_{x^n\sim P, y^m\sim Q}((\hat P_n,\hat Q_m)\in\Gamma({\mathcal{A}}))\nn\\
=&~\mathbf{P}_{x^n\sim P, y^m\sim Q}((\hat P_n^{\mathcal A},\hat Q_m^{\mathcal A})\in\Gamma^{\mathcal A} \cap\mathcal{P}_{n}(t)\times{\mathcal P_m}(t))\nn\\
\leq&~(n+1)^t(m+1)^t\max_{(R^\circ,S^\circ)\in\Gamma^{\mathcal A}\cap\mathcal{P}_{n}(t)\times{\mathcal P_m(t)}}\mathbf{P}_{x^n\sim P, y^m\sim Q}\left(\hat{P}_n=R^\circ,\hat{Q}_m=S^\circ\right)\nn\\
\leq&~(n+1)^t(m+1)^t \exp\left(-\inf_{(R,S)\in\Gamma}  \left(nD(R^{\mathcal A}\|P^\mathcal{A})+mD(S^{\mathcal A}\|Q^\mathcal{A})\right)\right),\nn
\end{align}
where the last two inequalities are from Lemmas~\ref{lem:numEmpDistribution} and \ref{lem:typeprob}. As the above holds for any $\mathcal A\in\Pi$, Eq.~(\ref{eqn:KLDdef}) indicates
\begin{align}
&~\limsup_{n\to\infty}\frac{1}{n+m}\log\mathbf{P}_{x^n\sim P, y^m\sim Q}((\hat P_n,\hat Q_m)\in\Gamma)\nn\\
\leq&~\inf_{\mathcal{A}}\left(-\inf_{(R,S)\in\Gamma}  \left(cD(R^{\mathcal A}\|P^\mathcal{A})+(1-c)D(S^{\mathcal A}\|Q^\mathcal{A})\right)\right)\nn\\
=&~-\sup_{\mathcal{A}}\inf_{(R,S)\in\Gamma} \left(cD(R^{\mathcal A}\|P^\mathcal{A})+(1-c)D(S^{\mathcal A}\|Q^\mathcal{A})\right).\nn
\end{align}
To obtain the lower bound, it remains to show  
\[\sup_{\mathcal{A}}\inf_{(R,S)\in\Gamma}\left(cD(R^{\mathcal A}\|P^\mathcal{A})+(1-c)D(S^{\mathcal A}\|Q^\mathcal{A})\right)\geq \inf_{(R,S) \in\operatorname{cl}\Gamma} \left(cD(R\|P)+(1-c)D(S\|Q)\right).\]




Assuming, without loss of generality, that the left hand side is finite, we only need to show

\[\operatorname{cl}\Gamma\cap B(P,Q,\eta)\neq\varnothing,\]
whenever \[\eta>\sup_{\mathcal{A}}\inf_{(R,S)\in\Gamma}\left(cD(R^{\mathcal A}\|P^\mathcal{A})+(1-c)D(S^{\mathcal A}\|Q^\mathcal{A})\right).\] Here $B(P,Q,\eta)$ is the divergence ball defined as follows
\[B(P,Q,\eta)=\left\{(R,S):cD(R\|P)+(1-c)D(S\|Q)\leq\eta\right\},\]
which is compact in $\mathcal P\times\mathcal P$~w.r.t.~the weak topology, due to the lower semi-continuity of $D(\cdot\|P)$ and $D(\cdot\|Q)$ as well as the fact that $0<c<1$.

To this end, we first show the following:
\begin{align}
\label{eqn:clGamma}
\operatorname{cl}\Gamma=\bigcap_{\mathcal{A}}\operatorname{cl}\Gamma(\mathcal{A}).
\end{align}
The inclusion is straightforward since $\Gamma\in\Gamma(\mathcal{A})$. The reverse means that if $(R,S)\in \operatorname{cl}\Gamma(\mathcal A)$ for each $\mathcal{A}$, then any neighborhood of $(R,S)$ w.r.t.~the weak convergence intersects $\Gamma$. To verify this, let $O(R,S)$ be a neighborhood of $(R,S)$ w.r.t.~the weak convergence, then there exists $\tilde{U}(R,S,\mathcal B,\zeta)\in O(R,S)$ over a finite partition $\mathcal B$ as $O(R,S)$ is also open in $\mathcal P_\tau\times\mathcal P_\tau$. Furthermore, the partition $\mathcal B$ can be chosen to refine $\mathcal A$ so that $\operatorname{cl}\Gamma(\mathcal B)\subset\operatorname{cl}\Gamma(\mathcal A)$. As $\tau$-topology is stronger than the weak topology,  a closed set in the $\mathcal P_\tau\times\mathcal P_\tau$ is closed in $\mathcal P\times\mathcal P$, and hence $\operatorname{cl}\Gamma(\mathcal B)\subset \operatorname{cl}_{\tau} \Gamma(\mathcal B)$. That $(R,S)\in\operatorname{cl}_\tau\Gamma(\mathcal B)$ implies that there exists $(R',S')\in\tilde{U}(R,S,\mathcal B,\zeta)\cap\Gamma(\mathcal B)$. By the definition of $\Gamma(\mathcal B)$, we can also find $(\tilde{R},\tilde{S})\in\Gamma$ such that $\tilde{R}(B_i)=R'(B_i)$ and $\tilde{S}(B_i)=S'(B_i)$ for each $B_i\in\mathcal B$, and hence  $(\tilde{R},\tilde{S})\in\tilde{U}(R,S,\mathcal B, \zeta)$. In summary, we have $(\tilde{R},\tilde{S})\in\tilde{U}(R,S,\mathcal B,\zeta)\subset O(R,S)$ and  $(\tilde{R},\tilde{S})\in\Gamma$. Therefore, $\Gamma\cap O(R,S)\neq\varnothing$ and the claim follows.

Next we show that, for each partition $\mathcal A$, 
\begin{align}
\Gamma(\mathcal A)\cap B(P,Q,\eta)\neq\varnothing.
\end{align}
By Eq.~(\ref{eqn:KLDdef}), there exists $(\tilde{P},\tilde Q)$ such that 
$cD(\tilde{P}^{\mathcal A}\|P^{\mathcal A})+(1-c)D(\tilde{Q}^{\mathcal A}\|Q^{\mathcal A})\leq\eta$. For such $(\tilde P, \tilde Q)$, we can construct $(P',Q')\in\Gamma(\mathcal A)$ as 
\begin{align}
P'(F)&=\sum_{i=1}^t\frac{\tilde{P}(A_i)}{P(A_i)}P(F\cap A_i),\nn\\
Q'(F)&=\sum_{i=1}^t\frac{\tilde{Q}(A_i)}{Q(A_i)}Q(F\cap A_i),\nn
\end{align}
for any measurable subset $F\subset\mathcal X$. If $P(A_i)=0$ ($Q(A_i)=0$) and hence $\tilde{P}(A_i)=0$ ($\tilde{Q}(A_i)=0$), as $D(\tilde P^{\mathcal A}\|P^{\mathcal A})<\infty$ ($D(\tilde Q^{\mathcal A}\|Q^{\mathcal A})<\infty$), for some $i$, the corresponding term in the above equation is set equal to $0$. Then $(P',Q')$ belongs to $\Gamma(\mathcal A)$ and also lies in $B(P,Q,\eta)$. The latter is because $D(P'\|P)=D(\tilde{P}^{\mathcal{A}}\|Q^{\mathcal A})$ and $D(Q'\|Q)=D(\tilde{Q}^{\mathcal A}\|Q^{\mathcal A})$: one can verify that any $\mathcal B$ that refines $\mathcal A$ satisfies \[D({P}'^{\mathcal B}\|P^{\mathcal B})=D(\tilde{P}^{\mathcal A}\|P^{\mathcal A}), D({Q}'^{\mathcal B}\|Q^{\mathcal B})=D(\tilde{Q}^{\mathcal A}\|Q^{\mathcal A}).\]

For any finite collection of partitions $\mathcal A_i\in\Pi$ and $\mathcal A\in\Pi$ refining each $\mathcal A_i$, each $\Gamma(\mathcal A_i)$ contains $\Gamma(\mathcal A)$. This implies that 
\[\bigcap_{i=1}^r\left(\Gamma(\mathcal A_i)\cap B(p,q,\eta)\right)\neq\varnothing,\]
for any finite $r$. Finally, the set $\operatorname{cl}\Gamma(\mathcal A)\cap B(P,Q,\eta)$ for any $\mathcal A$ is compact due to the compactness of $B(P,Q,\eta)$, and any finite collection of them has non-empty intersection. Thus, all these sets are also non-empty. This completes the proof.

\section{Proof of Theorem~\ref{thm:mainresult1}}
\label{sec:proofappendix}
	According to Theorem~\ref{thm:MMDmetrize}, $d_k$ metrizes weak convergence over $\mathcal P$. That $\alpha_{n,m}\leq\alpha$ can be verified by Lemma~\ref{lem:gammanm}, and we only need to show that the type-II error probability $\beta_{n,m}$ vanishes exponentially as $n$ and $m$ scale. For convenience, we will write the error exponent of $\beta_{n,m}$ as $\beta$.
	
	We first show $\beta\geq D^*$. With a fixed $\gamma>0$, we have $\gamma_{n,m}\leq \gamma$ for sufficiently large $n$ and $m$. Therefore,
	\begin{align}	
	\label{eqn:eqn1}
	\beta&=\liminf_{n,m\to\infty}-\frac{1}{n+m}\log \mathbf{P}_{x^n\sim P, y^m\sim Q}(d_k(\hat P_n, \hat Q_m)\leq\gamma_{n,m})\nn\\
	&\geq\liminf_{n,m\to\infty}-\frac{1}{n+m}\log\mathbf{P}_{x^n\sim P, y^m\sim Q}(d_k(\hat P_n, \hat Q_m)\leq\gamma)\nn\\
	&\geq\inf_{(R,S):d_k(R,S)\leq\gamma}\left(cD(R\|P)+(1-c)D(S\|Q)\right)\nn\\
	&\coloneqq D_\gamma^*,
	\end{align}
	where the last inequality is from the extended Sanov's theorem and that $d_k$ metrizes weak convergence of $\mathcal P$ so that $\{(R,S):d_k(R,S)\leq\gamma\}$ is closed in the product topology on $\mathcal P\times\mathcal P$. Since $\gamma>0$ can be arbitrarily small, we have 
	\[\beta\geq\lim_{\gamma\to 0^{+}}D^*_\gamma,\]
	where the limit on the right hand side must exist as $D^*_{\gamma}$ is positive, non-decreasing when $\gamma$ decreases, and bounded by $D^*$ that is assumed to be finite. Then it suffices to show 
	\begin{align}
	\lim_{\gamma\to0^{+}} D_\gamma^*=D^*.\nn
	\end{align}

	To this end, let $(R_\gamma, S_\gamma)$ be such that  $d_k(R_\gamma,S_\gamma)\leq\gamma$ and $cD(R_\gamma\|P)+(1-c)D(S_\gamma\|Q)=D^*_\gamma$. Notice that $R_\gamma$ and $S_\gamma$ must lie in
	\[\left\{W:D(W\|P)\leq\frac{D^*}c,  D(W\|Q)\leq\frac{D^*}{1-c}\right\}\coloneqq  \mathcal W,\]
	for otherwise $D_\gamma^*>D^*$. We remark that $\mathcal W$ is a compact set in $\mathcal P$ as a result of the lower semi-continuity of KLD w.r.t.~the weak topology on $\mathcal P$ \cite{VanErven2014Renyi,Dembo2009}. Existence of such a pair is a consequence of the facts that $\{(R,S):d_k(R,S)\leq \gamma\}$ is closed and convex, and that both $D(\cdot{\|P})$ and $D(\cdot\|Q)$ are convex functions \cite{VanErven2014Renyi}.
	
	Assume that $D^*$ cannot be achieved. We can write 
	\begin{align}
	\label{eqn:assu}
	\lim_{\gamma\to0^+}D^*_{\gamma}=D^*-\epsilon,
	\end{align}
	for some $\epsilon>0$. By the definition of lower semi-continuity, there exists a $\kappa_W>0$ for each $W\in\mathcal W$ such that 
	\begin{align}
	\label{eqn:geqhalf}
	cD(R\|P)+(1-c)D(S\|Q)\geq cD(W\|P)+(1-c)D(W\|Q)-\frac\epsilon 2	\geq  D^*-\frac\epsilon 2,
	\end{align}
	whenever $R$ and $S$ are both from 
	\[\mathcal S_W=\left\{R:d_k(R,W)<\kappa_W\right\}.\]
	Here the last inequality  comes from  the definition of $D^*$ given in Theorem~\ref{thm:mainresult1}. To find a contradiction, define 
	\[\mathcal S_W'=\left\{R:d_k(R,W)<\frac{\kappa_W}{2}\right\}.\]
	Since $S_W'$ is open and $\bigcup_W\mathcal S_{W}'$ covers $\mathcal W$, the compactness of $\mathcal W$ implies that there exists finite $\mathcal S_W'$'s, denoted by $\mathcal S_{W_1}',\ldots,\mathcal S_{W_N}'$, covering $\mathcal W$. Define $\kappa^*=\min_{i=1}^N\kappa_{W_i}>0$. Now let $\gamma<{\kappa^*}/{2}$ as $\gamma$ can be made arbitrarily small. Since $\bigcup_{i=1}^N \mathcal S'_{W_i}$ covers $\mathcal W$, we can find a $W_i$ with $R_\gamma\in \mathcal S_{W_i}'\subset\mathcal S_{W_i}$. Thus, \[d_k(S_\gamma,W_i)\leq d_k(S_\gamma,R_\gamma)+d_k(R_\gamma,W_i)<\kappa_{W_i}.\]
	That is, $S_\gamma$ also lies in $\mathcal S_{W_i}$. By Eq.~(\ref{eqn:geqhalf}) we get
	\[cD(R_\gamma\|P)+(1-c)D(S_\gamma\|Q)\geq D^*-\epsilon/2.\]
	However, by our assumption in Eq.~(\ref{eqn:assu}), it should hold that
	\[cD(R_\gamma\|P)+(1-c)D(S_\gamma\|Q)\leq D^*-\epsilon.\]
	Therefore, $\beta\geq D^*$.
	
	The other direction can be simply seen from the optimal type-II error exponent in Theorem~\ref{thm:upperbd}. Alternatively, we can use Chernoff-Stein lemma in a similar manner as in the proof of Theorem~\ref{thm:2sample1}. Let $P'$ be such that $cD(P'\|P)+(1-c)D(P'\|Q)=D^*.$
	Such $P'$ exists because $0<D^*<\infty$ and $D(\cdot\|P)$ and $D(\cdot\|Q)$ are convex w.r.t.~$\mathcal P$. That $D^*$ is bounded implies that both $D(P'\|P)$ and $D(P'\|Q)$ are finite. We  have 
	\begin{align}
	\beta_{n,m}=&~\mathbf{P}_{x^n\sim P, y^m\sim Q}(d_k(\hat P_n,\hat Q_m)\leq \gamma_{n,m})\nn\\
	\stackrel{(a)}{\geq}&~\mathbf{P}_{x^n\sim P, y^m\sim Q}(d_k(\hat P_n, P')+d_k(\hat Q_m, P')\leq{\gamma_{n,m}})\nn\\
	\stackrel{(b)}{\geq}&~\mathbf{P}_{x^n\sim P, y^m\sim Q}(d_k(\hat P_n, P')\leq\gamma_{n}, d_k(\hat Q_m, P')\leq{\gamma_{m}})\nn\\
	=&~\mathbf P_{x^n\sim P}(d_k(\hat P_n, P')\leq\gamma_{n})\,\mathbf P_{y^m\sim Q}(d_k(\hat Q_m, P')\leq{\gamma_{m}}),\nn
	\end{align}
	where $(a)$ and $(b)$ are from the triangle inequality of the metric $d_k$, and we pick $ \gamma_n=\sqrt{2K/n}(1+\sqrt{-\log\alpha})$, and  $\gamma_m=\sqrt{2K/m}(1+\sqrt{-\log\alpha})$ so that $\gamma_{n,m}>\gamma_n+\gamma_m$. Then Lemma~\ref{lem:gamman} implies  $\mathbf P_{x^n\sim P'}(d_k(\hat P_n,P')\leq\gamma_n)> 1-\alpha$. For now assume that $D(P'\|P)> 0$ and $D(P'\|Q)>0$. We can regard $\{x^n:d_k(\hat P_n,P')\leq\gamma_n\}$ as an acceptance region for testing $H_0:x^n\sim P'$ and $H_1:x^n\sim P$. Clearly, this test performs no better than the optimal level $\alpha$ test for this simple hypothesis testing in terms of the type-II error probability. Therefore, Chernoff-Stein lemma implies 
	\begin{align}
	\label{eqn:PP}
	\liminf_{n\to\infty}-\frac{1}{n}\log \mathbf P_{x^n\sim P}(d_k(\hat P_n,P')\leq\gamma_n)\leq D(P'\|P).
	\end{align}
	Analogously, we have
	\begin{align}
	\label{eqn:QQ}
	\liminf_{m\to\infty}-\frac{1}{m}\log \mathbf P_{y^m\sim Q}(d_k( \hat{Q}_m,P')\leq\gamma_m)\leq D(P'\|Q).
	\end{align}
	
	Now assume without loss of generality that $D(P'\|P)=0$, i.e., $P'=P$. Then $D(P'\|Q)>0$ under the alternative hypothesis $H_1:P\neq Q$, and Eq.~(\ref{eqn:QQ}) still holds. Using Lemma~\ref{lem:gamman}, we have $\mathbf P_{x^n\sim P}(d_k(\hat P_n,P')\leq\gamma_n)>1-\alpha$, which gives zero exponent. Therefore, Eq.~(\ref{eqn:PP}) holds with $P'=P$.
	
	As $\lim_{n,m\to\infty}\frac{n}{n+m}=c$, we conclude that	\[\beta=\liminf_{n,m\to\infty}-\frac{1}{n+m}\log\beta_{n,m}\leq D^*.\]
	
	The proof is complete.\QEDA

\section{Experiments: Kernel Choice vs.~Sample Size}
\label{sec:exp}
Following the discussion in Remark~\ref{rmk:kernelchoice}, we conduct a toy experiment to investigate how kernel choice and sample size affect the test type-II error probability. We consider Gaussian kernels that are determined by their bandwidth $\gamma$: $k(x,y)=\exp(-\|x-y\|^2/\gamma)$. The work \cite{Sutherland2017GeneandCrit} uses part of samples as training data to select the bandwidth, which we refer to the trained bandwidth in this paper. The estimated MMD is then computed using the trained bandwidth and the remaining samples. 

\begin{figure}[htbp]
	\centering
		\includegraphics[width=.5\linewidth]{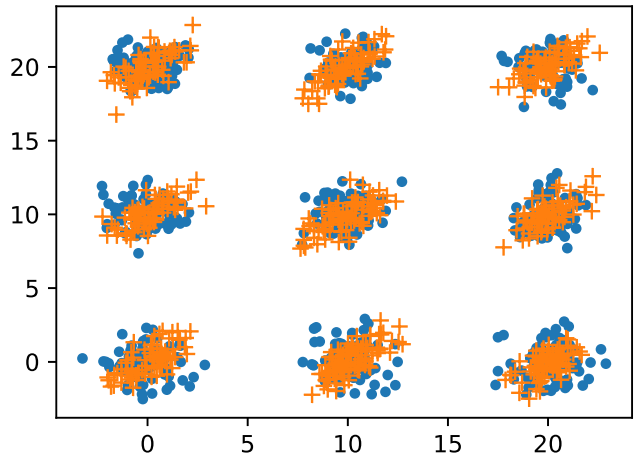}
		\caption{An example of samples drawn from distributions $P$ (blue dot) and $Q$ (orange plus sign).}
		\label{fig:data}
\end{figure}

\begin{figure}[htbp]
		\centering
		\includegraphics[width=0.55\linewidth]{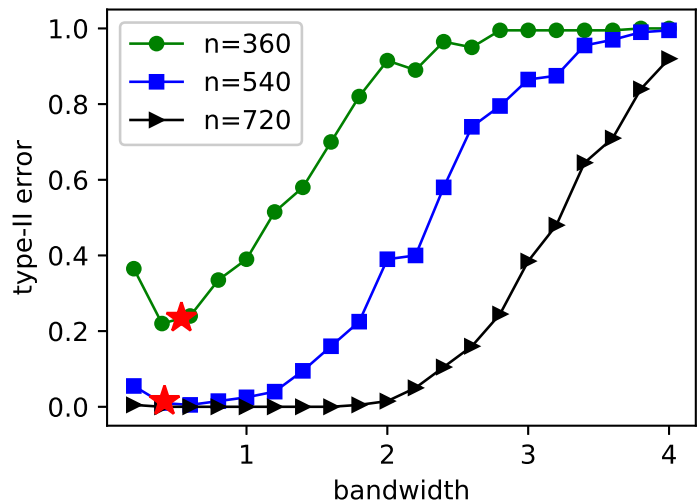}
		\caption{Experiment results for kernel choice vs. sample size. Red star denotes the trained bandwidth.}
		\label{fig:res1}
\end{figure}

We take a similar setting from \cite{Sutherland2017GeneandCrit} using the Blobs dataset \cite{Gretton2012OptKernelLarge}: $P$ is a $3\times 3$ grid of 2D standard Gaussians, with spacing $10$ between the neighboring centers; $Q$ is laid out identically, but with covariance $\frac{\epsilon-1}{\epsilon+1}$ between the
coordinates. Here we pick $\epsilon=6$ and generate $n=m=720$ samples from each distribution; an example of these samples is shown in Fig.~\ref{fig:data}. For our purpose, we pick splitting ratios $r=0.25$ and $r=0.5$ for computing the trained bandwidth. Correspondingly, there are $n=m=540$ and $n=m=360$ samples used to calculate the test statistic, respectively. With a level constraint $\alpha=0.1$, we report in Fig.~\ref{fig:res1} the type-II error probabilities over different bandwidths, averaged over $200$ trials, for each case with $n=m\in\{360, 540, 720\}$. The unbiased test statistic $d_u^2(\hat P_n,\hat Q_m)$ is used and the test threshold takes the minimum of the distribution-free threshold and the bootstrap one obtained by $500$ permutations \cite{Gretton2012}. We also mark the average trained bandwidths  corresponding to the respective sample sizes in the figure (red star marker). 


Fig.~\ref{fig:res1} verifies that the trained bandwidth is close to the optimal one in terms of the type-II error probability. Moreover, it indicates that a large range of bandwidths lead to lower or comparable error probabilities if we directly use the entire samples for testing. As the sample number increases, the exponential decay term in the type-II error probability becomes dominating and the effect of kernel choice diminishes. Since the desired range of bandwidths  is not known in advance, an interesting question is when we should split data for kernel selection and what is a proper splitting ratio.

\bibliographystyle{IEEEtran}	
\bibliography{SZhuBib}
 \end{document}